\documentclass[english,aps,prd,amssymb,superscriptaddress,nofootinbib,eqsecnum, twocolumn]{revtex4}
\usepackage[T1]{fontenc}
\usepackage[latin9]{inputenc}
\usepackage{amsmath}
\usepackage{amssymb}
\usepackage{graphicx}
\usepackage{esint}

\makeatletter

\DeclareRobustCommand{\greektext}{%
  \fontencoding{LGR}\selectfont\def\encodingdefault{LGR}}
\DeclareRobustCommand{\textgreek}[1]{\leavevmode{\greektext #1}}
\DeclareFontEncoding{LGR}{}{}
\DeclareTextSymbol{\~}{LGR}{126}
\providecommand{\tabularnewline}{\\}
\@ifundefined{textcolor}{}
{%
 \definecolor{BLACK}{gray}{0}
 \definecolor{WHITE}{gray}{1}
 \definecolor{RED}{rgb}{1,0,0}
 \definecolor{GREEN}{rgb}{0,1,0}
 \definecolor{BLUE}{rgb}{0,0,1}
 \definecolor{CYAN}{cmyk}{1,0,0,0}
 \definecolor{MAGENTA}{cmyk}{0,1,0,0}
 \definecolor{YELLOW}{cmyk}{0,0,1,0}
}

\def\ba{\begin{eqnarray}}
\def\ea{\end{eqnarray}}
\def\be{\begin{equation}}
\def\ee{\end{equation}}

\def\Msun{M_\odot}

\def\rhohat{{\hat\rho}}

\def\Cha{C_{h,\alpha}}

\def\Mha{M_{h,\alpha}}
\def\Rha{R_{h,\alpha}}
\def\eali{\eta_{\alpha,i}}
\def\eal{\eta_\alpha}
\def\gbarh{{\overline{g}_h}}

\def\nai{N_{\alpha, j}}
\def\nuai{\nu_{\alpha,j}}

\def\baj{b_{\alpha, i}}
\def\kai{\kappa_{(\alpha, i),j}}

\usepackage{babel}
\begin{document}

\title{Luminosity distance in Swiss cheese cosmology with randomized voids and galaxy halos}

%\author{E. E. Flanagan, N. Kumar, I. Wasserman}

\author{\'{E}anna \'{E}. Flanagan}

\email{flanagan@astro.cornell.edu}

\address{Laboratory for Elementary Particle Physics, Cornell University, Ithaca,
NY 14853, USA}

\address{Center for Radiophysics and Space Research, Cornell University, Ithaca,
NY 14853, USA}

\author{Naresh Kumar}

\email{nk236@cornell.edu}

\address{Laboratory for Elementary Particle Physics, Cornell University, Ithaca,
NY 14853, USA}

\author{Ira Wasserman}

\email{ira@astro.cornell.edu}

\address{Laboratory for Elementary Particle Physics, Cornell University, Ithaca,
NY 14853, USA}

\address{Center for Radiophysics and Space Research, Cornell University, Ithaca,
NY 14853, USA}

\begin{abstract}

We study the fluctuations in luminosity distance due to gravitational
lensing produced both by galaxy halos and large scale voids. Voids
are represented via a ``Swiss cheese'' model consisting of a $\Lambda$CDM
Friedman-Robertson-Walker background in which a number of randomly
distributed, spherical regions of comoving radius 35 Mpc are removed.
A fraction of the removed mass is then placed on the shells of the
spheres, in the form of randomly located halos, modeled with Navarro\textendash{}Frenk\textendash{}White
profiles. The remaining mass is placed in the interior of the spheres,
either smoothly distributed, or as randomly located halos. We compute
the distribution of magnitude shifts using a variant of the method
of Holz \& Wald (1998), which includes the effect of lensing shear.
In the two models we consider, the standard deviation of this distribution
is 0.065 and 0.072 magnitudes and the mean is -0.0010 and -0.0013 magnitudes,
for voids of radius 35 Mpc, sources at redshift 1.5, with the voids
chosen so that 90\% of the mass is on the shell today. The standard
deviation due to voids and halos is a factor $\sim3$ larger than
that due to 35 Mpc voids alone with a 1 Mpc shell thickness which we studied in our previous work. We also study the effect of the existence of evacuated voids,
by comparing to a model where all the halos are randomly distributed in the
interior of the sphere with none on its surface. This does not significantly
change the variance but does significantly change the demagnification
tail. To a good approximation, the variance of the distribution depends only on the mean column depth and concentration of halos and on the  
fraction of the mass density that is in the form of halos (as opposed to smoothly distributed):
it is independent of how the halos are distributed in space. We derive an approximate analytic formula for
the variance that agrees with our numerical results to $\lesssim 20\%$ out to $z\simeq 1.5$.
\end{abstract}
\maketitle

\section{Introduction}

\subsection{Background and Overview}

A number of surveys are being planned to determine luminosity distances
to various different astronomical sources, and to use them to constrain
properties of the dark energy or modification to gravity that drives
the cosmic acceleration. Perturbations to luminosity distances due
to gravitational lensing by large scale and galaxy scale structures
are a source of error for these studies, see, e.g., Refs \cite{p3Ref1, p3Ref2, p3Ref3, p3Ref4, p3Ref5, p3Ref6}. 

In this paper we use the computational method developed in Ref. \cite{p3Ref1}
to study the effect of density inhomogeneities on luminosity
distances in two idealized \textquotedblleft{}Swiss cheese\textquotedblright{}
models \cite{p3Ref3, p3Ref7, p3Ref8, p3Ref9, p3Ref10} of large scale ($\sim30$ Mpc) and galaxy scale
structures. Our models seek to capture the property that most of the
matter is concentrated in galaxy halos on the outer edges of voids
while the void interiors are relatively sparse. Our first model is
an extension of our previous work \cite{p3Ref1} where we idealize the interior
of a spherical void as a uniform underdense region and the surface of the sphere contains
a randomized distribution of galaxy halos with Navarro\textendash{}Frenk\textendash{}White
(NFW) profiles \cite{p3Ref13}. Our second model retains the randomly distributed
galaxy halos on the surface of the voids, and replaces the interior
uniform density with randomly distributed galaxy halos with NFW profiles.
In both models we keep fixed the parameters of the voids and halos.
Even though neither of these models represent realistic matter distributions
within a void, they should capture the main qualitative features
of lensing. 

\subsection{Summary of Computation and Results}

In Section II, we give a detailed description of the NFW halo density
profile that we adopt which is motivated by observations. We also
describe our two models of the entire distribution of matter, including
both large scale void structures and smaller scale halo structures. 

In Section III, we review the method we use to compute the distribution of
lensing magnifications. We then describe how to compute the lensing convergence
for a single halo. We compare our numerical results for the distribution
of magnifications with analytic expressions. We also estimate
the number of realizations required to get a reasonable accuracy in
the computed distribution of magnifications (e.g, obtain the mean
of the distribution to an accuracy of $<1\%$ ). The accuracy of the
numerical results scales as $N^{-1/2}$ as expected, where $N$ is
the number of realizations.

In Section IV, we study our first Swiss cheese model. We start by
describing the model, and study the propagation of light rays through just a single
void. Then, we derive analytic results for the magnification
and use these to check our numerical results. We study the expected 
number of halo intersections, the redshift dependence of the magnification distribution,
and determine the contribution of shear to our results. We note that the standard deviation is $\sim 3$ times larger than that due to voids with no halos, specifically the model of  \cite{p3Ref1} consisting of voids of radius 35 Mpc with smooth underdense interiors and a smooth overdensity concentrated on the surface of the sphere with a thickness of 1 Mpc. We show that the redshift
dependence of the mean and standard deviation agrees with analytic
results to $<10\%$. We also note that the standard deviation changes by less than $3\%$
if shear is neglected (see Section IV C below).

One effect which our models do not include is the clustering of halos,
that is, the correlations between the locations of different halos.
While it would be more realistic to include the effects of clustering,
our simplified models should capture the essence of the effects of
large scale inhomogeneities.

In Section V, we study the second Swiss cheese model. Again, we first
describe the model and study just a single void. We derive analytic
results for the lensing convergences and use these to check our numerical
results. There is a higher probability
of demagnification; this shift is expected because the density contrast
inside the void is now sharper because it is empty (with a smattering
of a small number of halos) whereas the first model has a smooth
interior matter distribution. The redshift dependence of the mean
and standard deviation of the second model are similar to those of
the first model. In the two models we consider, the standard deviations
of this distribution are 0.065 and 0.072 magnitudes and the means are -0.0010
and -0.0013 magnitudes, for voids of radius 35 Mpc, sources at redshift
1.5, with the voids chosen so that 90\% of the mass is on the shell
today. We compare the
distributions for configurations with and without voids for a source
at $z_{s}=1.5$. We find that the voids do not significantly change
the variance but do significantly change the demagnification tail
and the mode.

We find that since the distribution is skewed, the mode is positive, while the variance is determined primarily by rays that intersect halos. The scale of the voids does not significantly influence our results. The main parameters that determine the mode and variance of the distribution is the mean column depth and concentration of halos and  the fraction of the mass density that is in the form of halos (as opposed to smoothly distributed). The distribution of halos in space (i.e., in the interior versus the surface) is unimportant. Hence, our models bracket the range of possibilities of magnifications. Our analysis is generally consistent with other analytic and computational
results \cite{p3Ref14, p3Ref15, p3Ref16, p3Ref17, p3Ref18, p3Ref19, p3Ref20, p3Ref21, p3Ref22}. We also compare our results to those of Kainulainen \& Marra
\cite{p3Ref11, p3Ref12} who use a similar but slightly different simplified model
of large scale structure.

\section{Model of Lensing Due to Galaxy halos and Voids}

\subsection{Galaxy halo profile}

We model the galaxy halos with an NFW profile \cite{p3Ref13}, with a density
distribution

\begin{equation}
\rho_{\mathrm{halo}}\left(r\right)=\left\{ \begin{array}{c}
0\;\;\;\;\;\;\;\;\;\;\;\;\;\; r\geqslant CR_{s}\\
\frac{\rho_{0}R_{{\rm s}}^{3}}{r\left(r+R_{{\rm s}}\right)^{2}}\;\;\; r\leqslant CR_{s}
\end{array}\right..\label{3p2.1}
\end{equation}
Here $r$ is the proper spherical radial coordinate, $R_{s}$ is the physical
radius which defines the core of the halo where most of the mass is
concentrated, $C$ the ratio of the radius of the halo to the core
radius $R_{s}$, and the parameter $\rho_{0}$ is determined by
the total mass of the halo. The corresponding total halo mass is
\begin{equation}
M_{\mathrm{halo}}=4\pi\rho_{0}R_{{\rm s}}^{3}\left(\log\left(1+C\right)-\frac{C}{1+C}\right).\label{3p2.2}
\end{equation}
For all our simulations we use $M_{\mathrm{halo}} = 1.25\times10^{12}M_{\odot}$,
$R_{s}=30\;\mathrm{kpc}$ and $C=10$ \cite{MW1, MW2, MW3}. These values determine the
halo density parameter $\rho_{0}$. This completely defines our NFW
halo model and we list our parameters in Table I. In this paper we
keep the halo parameters fixed, but it would be straightforward to
explore other values.

\begin{table}
\begin{tabular}{|c|c|}
\hline 
\multicolumn{1}{|c||}{Quantity} & Value\tabularnewline
\hline 
\hline 
$M_{\mathrm{halo}}$ & $1.25\times10^{12}M_{\odot}$ \tabularnewline
\hline 
$R_{{\rm s}}$ & 0.03 Mpc\tabularnewline
\hline 
$C$ & 10\tabularnewline
\hline 
\end{tabular}

\caption{Parameters of halo with NFW profile}
\end{table}

\subsection{Our void models}

In Swiss cheese models, the Universe contains a network of spherical,
non-overlapping, mass-compensated voids. The voids are chosen to be
mass compensated so that the potential perturbation vanishes outside
each void. Mass flows outward from the evacuated interior and is then
trapped on the shell wall. In our previous work, \cite{p3Ref1}, we considered
a uniformly underdense interior with a $\delta$-function shell on
the surface. This model is determined by a fixed comoving radius $R$
and by the fraction, $f$, of the total void mass on the shell today.
These parameters determine the evolution with time of the interior
mass density and the surface mass density.

In this paper we generalize the models of \cite{p3Ref1} to include the halo
substructure of the voids. We consider two different idealized models.
In the first, each void consists of a central, uniformly
underdense region surrounded by a shell consisting of randomly distributed
halos, and in the second, halos are placed randomly both in
the interior and on the surface. The zero thickness shell is thus
replaced by halos randomly distributed on the surface of the sphere,
with the number of halos chosen to match the mass of the shell. The
number of halos thus evolves with time. We call our first model the
Swiss Raisin Nougat (SRN) model, with ``raisins'' denoting halos
and ``nougat'' the smooth void interior. We call the second model
the Swiss Raisin Raisin (SRR) model. 

For a given void, we denote by $\mathbf{r}$ the physical displacement from the center of the void at $\mathbf{r} = 0$, and we denote by $\mathbf{s} = \mathbf{r} a_{\mathrm{ex}}\left(z\right)$ the comoving displacement, where $a_{\mathrm{ex}}\left(z\right)$ is the scale factor of the background $\Lambda$CDM Friedman-Robertson-Walker (FRW) cosmology. The quantity that determines the lensing magnification is the density perturbation 
\begin{equation}
\Delta\rho\left(\mathbf{r},\; z\right)=\rho\left(\mathbf{r},\; z\right)-\rho_{{\rm FRW}}\left(z\right),\label{3p2.3}
\end{equation}
where $\rho_{{\rm FRW}}\left(z\right)=3H_{0}^{2}\Omega_{{\rm M}}/\left(8\pi Ga_{{\rm ex}}^{3}\left(z\right)\right)$ is the background FRW density and $z$ is redshift. For the SRN model the density perturbation is 
\[
\Delta\rho_{\rm{SRN}}\left(\mathbf{r}\right)=-f\left(z\right)\rho_{\mathrm{FRW}}\left(z\right)\Theta\left(a_{{\rm ex}}Y_{{\rm void}}-r\right)
\]
\begin{equation}
+\sum_{i=1}^{N_{{\rm shell}}\left(z\right)}\rho_{{\rm halo}}\left(\left|\mathbf{r}-a_{{\rm ex}}Y_{{\rm void}}\mathbf{\hat{n}}_{i}\right|\right),\label{3p2.4}
\end{equation}
where the first term is the smoothly distributed interior underdensity and the second term is due to halos on the surface. Here $f\left(z\right)$ is the fraction of the mass of the sphere on the surface \cite{p3Ref1}, $Y_{\mathrm{void}}$ is the (constant) comoving void radius, $\Theta$ is the step function, $N_{\mathrm{shell}}$ is the number of halos on the surface, and $\mathbf{\hat{n}}_{i}$ is a randomly chosen unit vector giving the location of the $i$-th halo on the surface of the sphere. The number of surface halos is 
\begin{equation}
N_{{\rm shell}}\left(z\right)=f\left(z\right)\frac{M_{{\rm void}}}{M_{{\rm halo}}},\label{3p2.5}
\end{equation}
where 
\begin{equation}
M_{{\rm void}}=\frac{4}{3}\pi Y_{{\rm void}}^{3}a_{{\rm ex}}^{3}\rho_{{\rm FRW}}\label{3p2.6}
\end{equation}
is the conserved total void mass.

For the SRR model, the density perturbation is 
\[
\Delta\rho_{{\rm SRR}}\left(\mathbf{r}\right)=-\rho_{{\rm FRW}}\left(z\right)\Theta\left(a_{{\rm ex}}Y_{{\rm void}}-r\right)
\]
\[
+\sum_{i=1}^{N_{{\rm shell}}\left(z\right)}\rho_{{\rm halo}}\left(\left|\mathbf{r}-a_{{\rm ex}}Y_{{\rm void}}\mathbf{\hat{n}}_{i}\right|\right)
\]
\begin{equation}
+\sum_{i=1}^{N_{{\rm core}}\left(z\right)}\rho_{{\rm halo}}\left(\left|\mathbf{r}-a_{{\rm ex}}Y_{{\rm void}}\mathbf{m}_{i}\right|\right),\label{3p2.7}
\end{equation}
where the last term represents the halos in the interior. Here $N_{\mathrm{core}}\left(z\right) = \left(1-f\left(z\right)\right) M_{\mathrm{void}}/M_{\mathrm{halo}}$ is the number of interior halos and the vectors $\mathbf{m}_j$ are randomly chosen in the interior of the unit sphere.

Now consider a light ray that intersects the void. A key role in our computations will be played by the impact parameters of the ray with respect to the center of the void, and with respect to the centers of the halos. These impact parameters will be two dimensional vectors in the plane perpendicular to the unperturbed ray. Specifically, we introduce a basis of three orthonormal spatial vectors $\mathbf{e}_1$, $\mathbf{e}_2$ and $\mathbf{e}_3$ with $\mathbf{e}_3$ along the direction of the ray. We denote by 
\begin{equation}
\mathbf{p}=\sum_{A=1,\;2}p^{A}\mathbf{e}_{A}\label{3p2.8}
\end{equation}
the comoving impact parameter of the ray with respect to the center of the void. We denote by
\begin{equation}
\mathbf{b}_{i}=\sum_{A}b_{i}^{A}\mathbf{e}_{A}=\sum_{A}a_{{\rm ex}}\left[p^{A}-Y_{{\rm void}}\hat{n}_{i}^{A}\right]\mathbf{e}_{A}\label{3p2.9}
\end{equation}
the physical impact parameter of the ray with respect to the center of the $i$-th halo on the surface, where we have decomposed the unit vectors $\mathbf{\hat{n}}_{i}$ as $\mathbf{\hat{n}}_{i}=\sum_{A}\hat{n}_{i}^{A}\mathbf{e}_{A}+\hat{n}_{i}^{3}\mathbf{e}_{3}.$ Similar formulae are obtained for the impact parameters of the interior halos.

Even though the SRN and SRR models are highly idealized, they are more realistic
than the void models in our previous work \cite{p3Ref1}. A key feature of
our idealized models is that they can be evolved in time continuously
and very simply. Within the context of this highly idealized class
of models, we study the distribution of magnitude shifts relative
to what would be found in a smooth cold dark matter (CDM) model of
the Universe with a cosmological constant, \textgreek{L}, for different
source redshifts. 

It is important to note that our models are not spherically symmetric,
as we break up the shell to form halos. We assume that nevertheless
the large scale evolution of a void is the same as it would be in
spherical symmetry. We also neglect gravitational clustering of halos
on void surfaces. Our main aim is to investigate the role of small
scale clumps in producing magnitude shifts. 

To compute the effects of rays passing through our cosmology, we follow
the steps described in Section IIC of \cite{p3Ref1}, with the added halo
contributions. Specifically, we compute a $4\times4$ matrix for each
void, multiply all the matrices together, and compute the total magnification
from the final $4\times4$ matrix. The explicit expressions for the
$4\times4$ matrices in term of line integrals of derivatives of the
gravitational potential are given in Eqs. (2.18) - (2.20) of \cite{p3Ref1}.
We drop all of the integrals over the projected Riemann tensor in
Eqs. (2.19) of \cite{p3Ref1} except the one in the formula for $L_{\: C}^{A}$. We then repeat the computation $N>>1$ times to build up the distribution of magnifications.

\section{Results for a Single Halo}

\begin{figure}
\includegraphics[scale=0.46]
{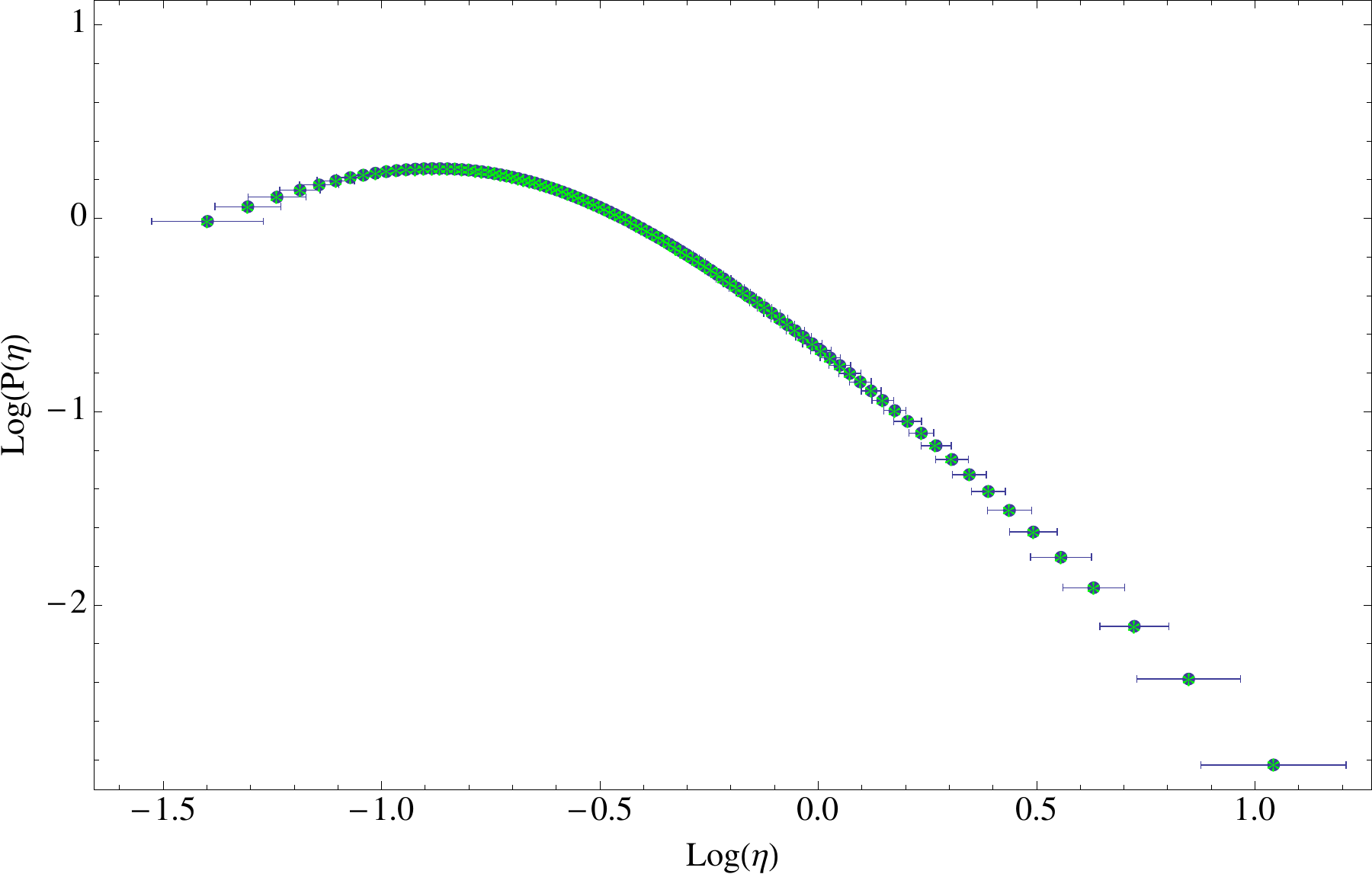}
\centering
\caption{Comparison between numerical (points with ranges) and analytic results
(starred points) for the distribution of integrated column depths.}
\end{figure}

We now discuss the distribution of magnifications due to a single
halo. The halo has two distinct regions, the core $x<R_{s}$ and the
external region $R_{s}<x<CR_{s}$. Here $x$ is the physical distance.
To compute the magnification, we first compute the lensing convergence,
$\kappa$, analytically. For a general density contrast $\delta\left(\mathbf{s}\right) = \Delta \rho\left(\mathbf{s}\right)/\rho_{\mathrm{FRW}}$ this is given by 
\begin{equation}
\kappa=\frac{3}{2}\frac{H_{0}^{2}}{c^{2}}\Omega_{{\rm M}}\int_{0}^{y_{S}}dy\frac{y\left(y_{S}-y\right)}{y_{S}a_{{\rm ex}}\left(z\right)}\delta\left(y,\; z\right),\label{3p3.1}
\end{equation}
where $y$ is comoving distance along the ray, $y_{S}$ is the comoving distance
to the source, $a_{{\rm ex}}\left(z\right)=\left(1+z\right)^{-1}$,
$H_{0}$ is the Hubble constant, $c$ is the velocity of light, $\Omega_{{\rm M}}$
is the matter fraction and $z=z\left(y\right)$ is redshift. Combining the halo profile (\ref{3p2.1}) and the second term in Eq. (\ref{3p2.4}) with Eq. (\ref{3p3.1}) gives for the lensing convergence due to the halo
\[
\kappa\left(b\right)=\frac{8\pi Ga_{{\rm ex}}\left(z\right)}{c^{2}}(\rho_{0}R_{s})\frac{y\left(y_{{\rm S}}-y\right)}{y_{{\rm S}}}
\]
\begin{equation}
\left[\kappa_{{\rm core}}\Theta\left(R_{s}-b\right)+\kappa_{{\rm out}}\Theta\left(b-R_{s}\right)\Theta\left(CR_{s}-b\right)\right].
\label{3p3.3}
\end{equation}
Here $b=\left|\mathbf{b}\right|$ is the physical impact parameter
\[
%\begin{eqnarray}
\kappa_{{\rm core}}=\left(-\frac{\sqrt{C^{2}-\alpha}}{\left(1-\alpha\right)\left(1+C\right)}+\frac{2}{\left(1-\alpha\right)^{3/2}}\right.
\]
\begin{equation}
\left.\left[\tanh^{-1}\left(\frac{\sqrt{1-\sqrt{\alpha}}}{\sqrt{1+\sqrt{\alpha}}}\right)-\tanh^{-1}\left(\frac{\sqrt{1-\alpha}}{C+\sqrt{C^{2}-\alpha}+1}\right)\right]\right)
\label{3p3.4}
\end{equation}
%\end{eqnarray}
and

\[
\kappa_{{\rm out}}=
\left(\frac{\sqrt{C^{2}-\alpha}}{\left(\alpha-1\right)\left(1+C\right)}-\right.\frac{2}{\left(\alpha-1\right)^{3/2}}
\]

\begin{equation}
\left.\left[\tan^{-1}\left(\frac{\sqrt{\sqrt{\alpha}-1}}{\sqrt{\sqrt{\alpha}+1}}\right)-\tan^{-1}\left(\frac{\sqrt{\alpha-1}}{C+\sqrt{C^{2}-\alpha}+1}\right)\right]\right).\label{3p3.5}
\end{equation}
Here $\Theta$ is the step function and $\alpha=b^{2}/R_{{\rm s}}^{2}$.
Note that $\alpha<1$ for $b<R_{{\rm s}}$, $\alpha>1$ for $R_{{\rm s}}<b<CR_{{\rm s}}$$ $
and $\kappa=0$ for $b>CR_{{\rm s}}$.

Finally, the mean of the lensing convergence for a single halo is
obtained by averaging over the impact parameter
\begin{equation}
\hat{\kappa}=\frac{2}{\left(CR_{{\rm s}}\right)^{2}}\int_{0}^{CR_{{\rm s}}}\kappa\left(b\right)bdb\label{3p3.6}
\end{equation}
\begin{equation}
\;\;\;\;\;\;\;\;\;\;\;\;\;\;\;\;\;\;\;\;\;\hat{\kappa}=\frac{8\pi Ga_{{\rm ex}}\left(z\right)}{c^{2}}(\rho_{0}R_{s})\frac{y\left(y_{{\rm S}}-y\right)}{y_{{\rm S}}}\frac{M_{{\rm halo}}}{\left(CR_{{\rm s}}\right)^{2}}.\label{3p3.7}
\end{equation}

\begin{figure}
\includegraphics[scale=0.55]{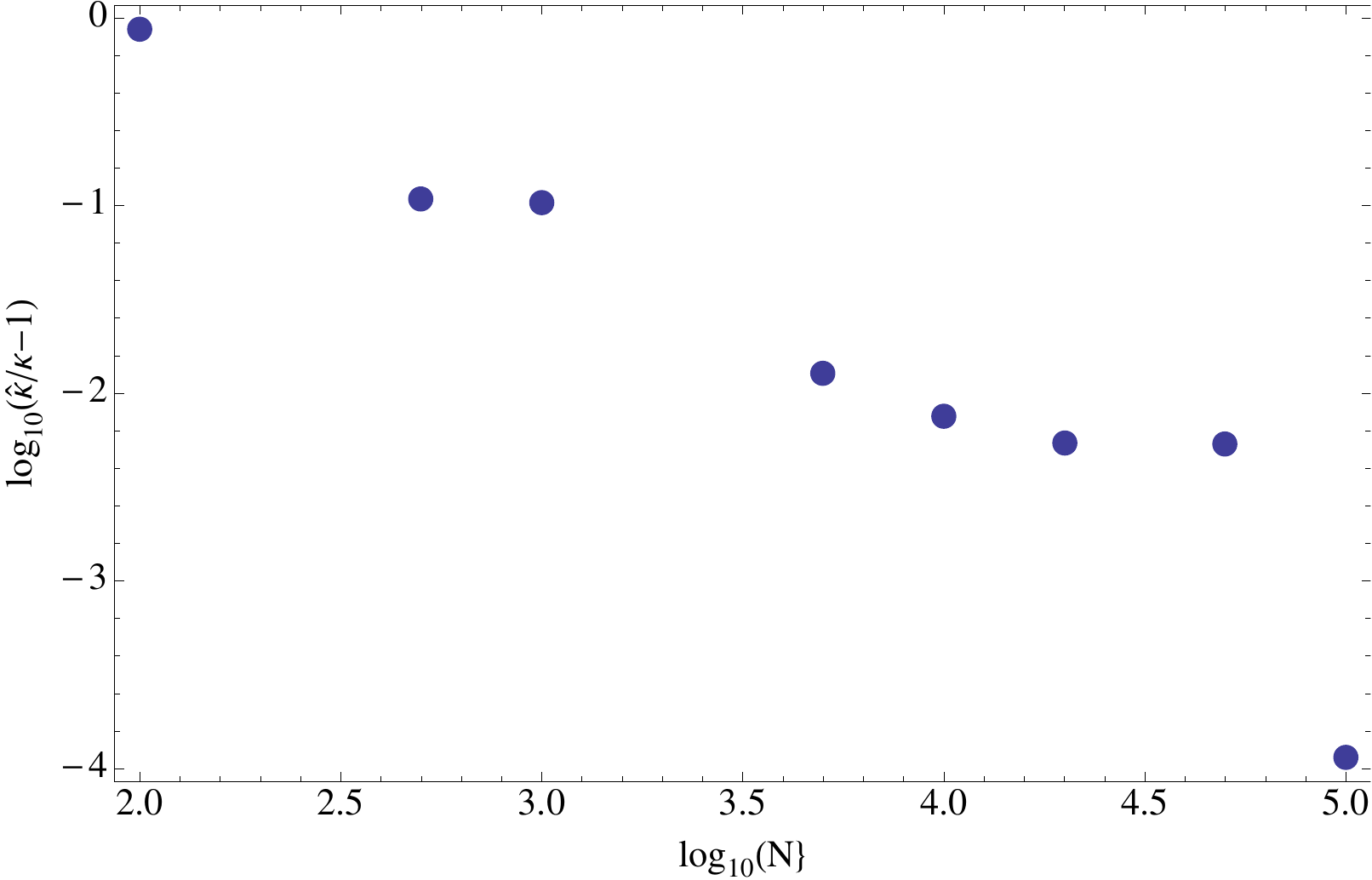}

\caption{The difference between numerically $\left(\kappa\right)$ and analytically
computed mean convergences $\left(\hat{\kappa}\right)$ as a function
of number of runs $N$, for one void, comoving void radius $R$ =
35 Mpc, and fraction of void mass on the shell today $0.9$. Our numerical
simulations agree with the analytic result to $<1\%$ for $N=10^{4}$.}
\end{figure}

We define $\eta=\int\rho\left(z\right)dz$ to be the integrated column
density which is proportional to the convergence $\kappa$. Figure
1 is a comparison of $\log_{10}\left(P\left(\eta\right)\right)$,
the logarithm of the probability distribution of $\eta$, computed
analytically (starred points, Eq. (\ref{3pA5}) from Appendix A) and the results
from our code (points associated with the $\eta$ bins). Within each
$\eta$ bin, the mean of the bins agree well with analytic results.
The width of the bins represents the sampling accuracy within those
bins, the centers of halos are sampled less than the rest of the
halos.

To further assess the accuracy of our numerical results, we compute
the mean of the distribution for a single halo for different numbers
of runs ($N$), and compare this with the analytic expression (\ref{3p3.7}). We find that the results from our numerics agree with
the theoretical prediction with an accuracy $\sim N^{-1/2}$ as expected.
In Figure 2, we plot the estimator of the mean as a function of $N$.
We see that the accuracy is <1\% for $N=10^{4}$. For the rest of the simulations in this paper,
we will use $N=10^{4}$.

\section{Results for the Swiss Raisin Nougat Model}

The Swiss Raisin Nougat (SRN) model is an idealized Swiss Cheese model
containing spherical voids with comoving radius $Y_{{\rm void}}=35$
Mpc. As explained earlier, the matter in the interior moves towards
the outer edges of the void with the evolution of the Universe. For
a particular void at some redshift, we break up the mass on the shell
of the void into halos with NFW profiles and randomly distribute them
on the shell. The mass in the interior is smeared smoothly inside
the sphere with a uniform mass density. The parameters of this model
are listed in Table II.

A key change from our void models in \cite{p3Ref1} is that there is no longer
a zero thickness shell. One of the issues encountered in that model
was the logarithmic divergence in the variance of the lensing convergence
distribution due to the zero thickness assumption. Here, however,
we break up the void surface into halos and the effective thickness
of the shell is set by the size of these halos which acts as a natural
cutoff. Hence, the divergence is avoided which makes for a more realistic
and robust model. 

\begin{table}
\begin{tabular}{|c|c|}
\hline 
\multicolumn{1}{|c||}{Quantity} & Value\tabularnewline
\hline 
\hline 
$\Omega_{{\rm M}}$ & 0.3\tabularnewline
\hline 
$\Omega_{\Lambda}$ & 0.7\tabularnewline
\hline 
$H_{0}$ & 70 $\mathrm{kms^{-1}Mpc^{-1}}$\tabularnewline
\hline 
$Y_{{\rm void}}$ & $35$ Mpc\tabularnewline
\hline 
Halo profile & NFW\tabularnewline
\hline 
Present fraction of void mass on shell & 0.9\tabularnewline
\hline 
Fraction of shell mass in halos & 1.0\tabularnewline
\hline 
Fraction of interior mass in halos & 0.0\tabularnewline
\hline 
\end{tabular}

\caption{Parameters of SRN model}
\end{table}

\subsection{Probability of intersecting a halo}

The expected number of times a light ray hits a halo is given by the
ratio of the total projected area of all the halos in a void to the
projected area of the void. The expected number of intersections at
comoving impact parameter $p = \left|\mathbf{p}\right|$ (comoving distance from the center
of the void) through the shell at redshift $z$ is
\[
N_{{\rm int}}\left(p,\: z\right)=\frac{f\left(z\right) M_{{\rm void}}\left(z\right)}{4\pi Y_{{\rm void}}^{2}a_{{\rm ex}}^{3}\left(z\right)M_{{\rm halo}}}\pi R_{{\rm halo}}^{2}a_{{\rm ex}}^{3}\left(z\right)\times
\]

\begin{equation}
\int_{0}^{\sqrt{Y_{{\rm void}}^{2}-p^{2}}}ds\delta\left(\sqrt{s^{2}+p^{2}}-Y_{{\rm void}}\right)\label{3p4.1}
\end{equation}
\begin{equation}
=\frac{f\left(z\right) M_{{\rm void}}\left(z\right)R_{{\rm halo}}^{2}}{2Y_{{\rm void}}\sqrt{Y_{{\rm void}}^{2}-p^{2}}a_{{\rm ex}}^{2}\left(z\right)M_{{\rm halo}}}.\label{3p4.2}
\end{equation}
Here $s$ is the comoving distance from the center of a void and $R_{{\rm halo}}=CR_{{\rm s}}$ is the
physical radius of the halo. Averaging over the impact parameter, $p$, gives
\begin{equation}
N_{{\rm int}}\left(z\right)=\frac{f\left(z\right) M_{{\rm void}}\left(z\right)R_{{\rm halo}}^{2}}{Y_{{\rm void}}^{2}a_{{\rm ex}}^{2}\left(z\right)M_{{\rm halo}}}.\label{3p4.3}
\end{equation}
Note that the void radius $Y_{{\rm void}}$ is comoving while the
halo radius $R_{{\rm halo}}$ is physical. Both these parameters are
fixed and do not evolve with time. For example, for a void placed at redshift
0, $N_{{\rm shell}}\left(0\right)\simeq0.4$.

\begin{figure}
{\includegraphics[scale=0.44]{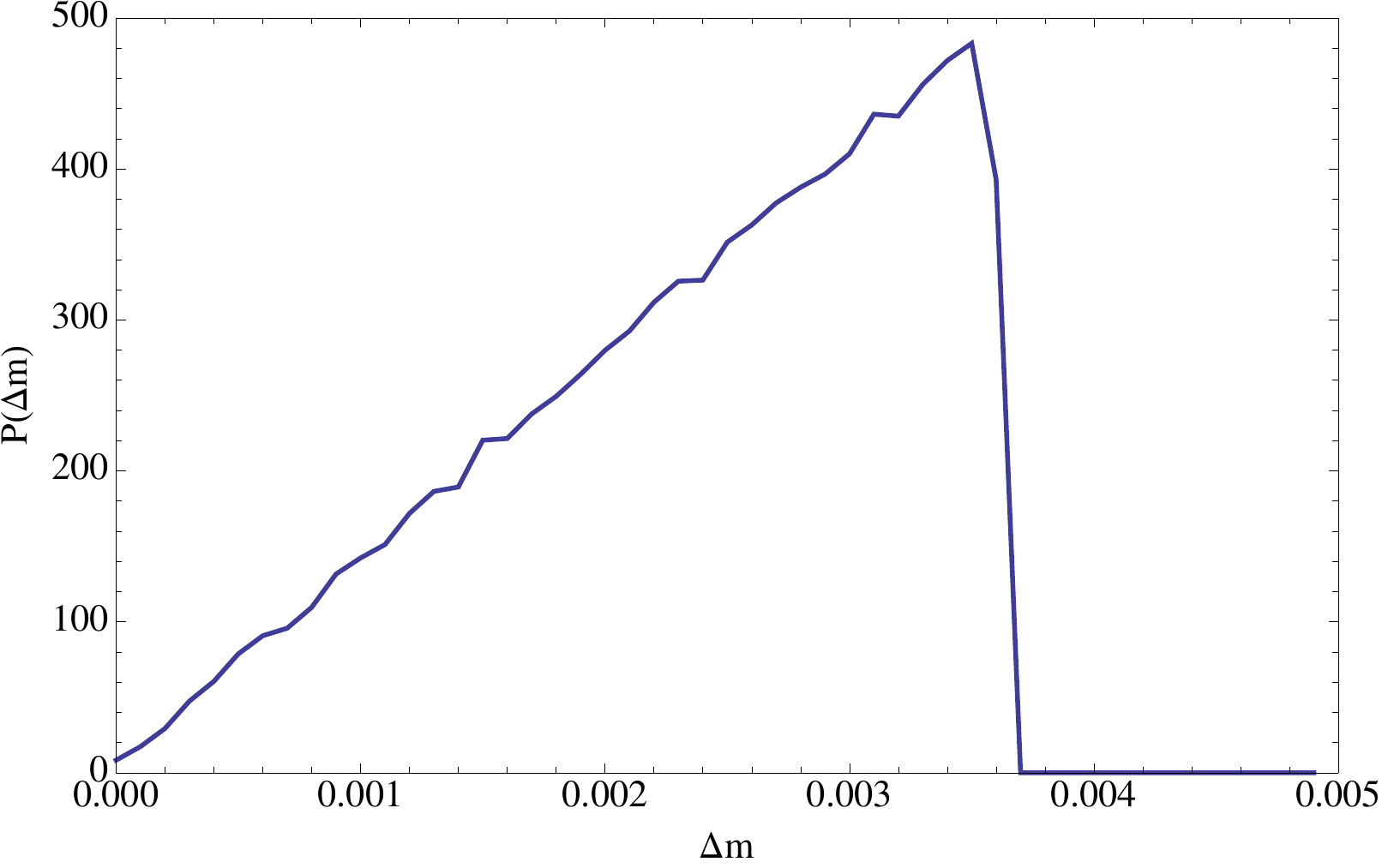}

}

{\includegraphics[scale=0.5]{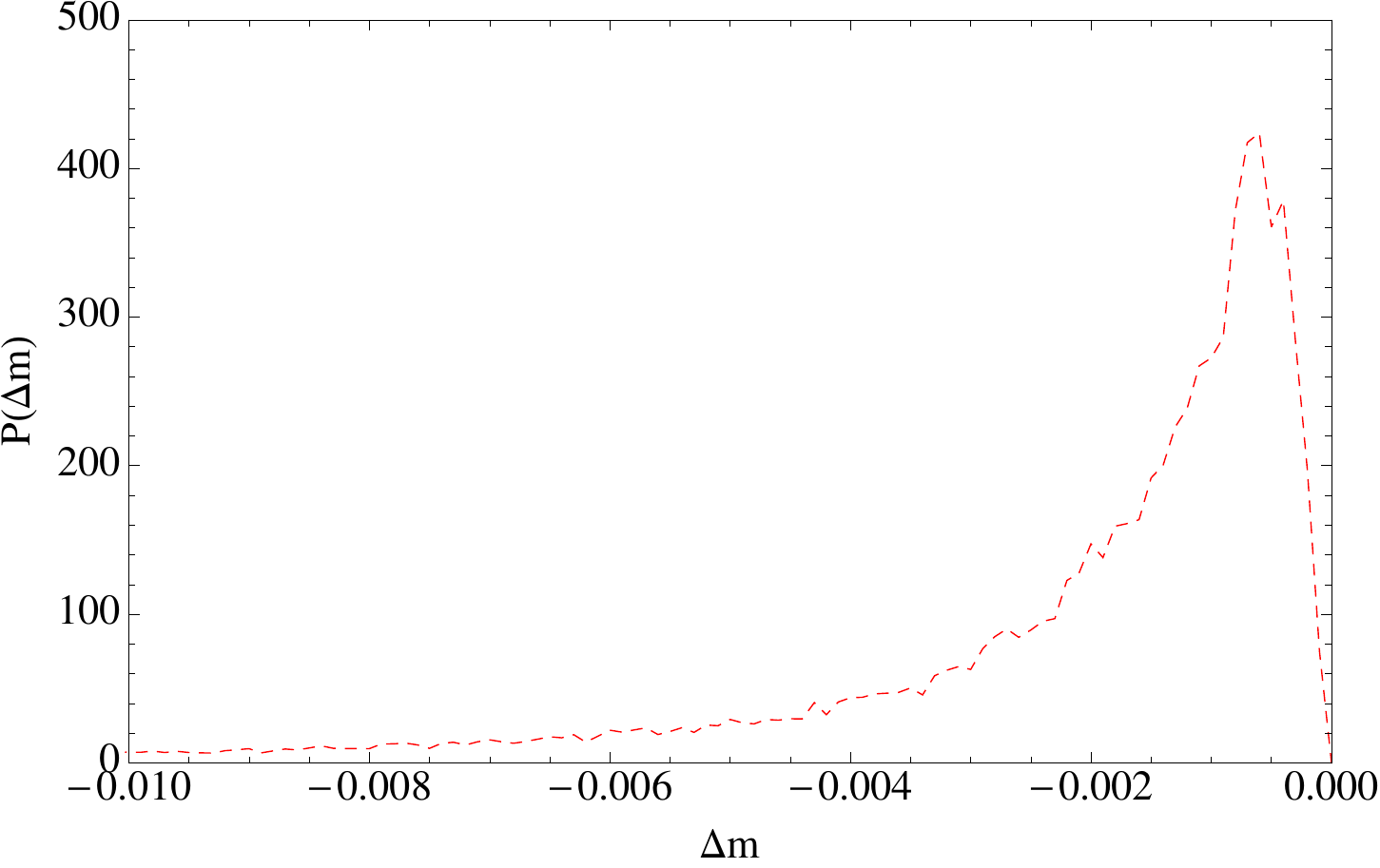}

}

\caption{The probability distribution of magnitude shifts $\Delta m$ for a
simulation in the SRN model with $\Omega_{M}=0.3$, with one void
of radius 35 Mpc at $z=0.45$ and sources placed at $z_{s}=1$, and
fraction of void mass on the shell today $f=0.9$. Top: The probability distribution of magnitude shifts $\Delta m$ for a single void for $\Delta m$ positive. Bottom: The probability distribution for magnitude shifts $\Delta m$ for a single void for $\Delta m$ negative. The total probability
for $\Delta m<0$ is $\simeq0.8.$}
\end{figure}

\subsection{One void}

In this section we will focus on a single void at redshift
0.45 and a source at redshift 1. We calculate the expected number
of halo intersections for a light ray from Eq. (\ref{3p4.3}). For a 35 Mpc
void, using the halo parameters from Section II, we get $N_{{\rm shell}}\left(0.45\right)\simeq0.8$.
For $N=10^{4}$ runs, we keep track of the number of times that light
rays hit halos and we obtain 8064 instances, which agrees well with
our prediction. We use the density perturbation (\ref{3p2.4}) for the SRN
model and compute the lensing convergence $\kappa$ by summing the result (\ref{3p3.3}) over all intersected halos.

For the rest of the paper, we will concentrate on the distribution
of the magnitude shift $\Delta m$, which is a function of lensing
convergence $\kappa$
\begin{equation}
\Delta m=\frac{2.5}{\ln10}\ln\left|\left(1-\kappa\right)^{2}-\gamma^{2}\right|.\label{3p4.4}
\end{equation}
Here $\gamma$ is the shear, which we will discuss in the next section.
Figure 3 shows the probability distribution $P\left(\Delta m\right)$ of magnitude shifts $\Delta m$ we obtain
for a single void placed at $z=0.45$ (without shear). 

A notable feature of this distribution is that it is bimodal, with
peaks at both positive and negative $\Delta m$ (We plot separately
the distribution for positive $\Delta m$ and for negative $\Delta m$,
since the relevant scales for these two regions of the probability
distribution are very different). The peak at positive $\Delta m$
is predominantly due to rays that do not intersect any halos, and
are demagnified by their passage through the underdense void interior.
The peak at negative $\Delta m$ is predominantly due to rays which
intersect one or more halos and are consequently magnified. 

We can also compute the mean of each of these distributions and compare
them with analytical expressions. The means for $\Delta m>0$, $\left\langle \Delta m\right\rangle _{+}$
and for $\Delta m<0$, $\left\langle \Delta m\right\rangle _{-}$
are defined to be
\begin{equation}
\left\langle \Delta m\right\rangle _{+}=\int_{0}^{\infty}\Delta mP\left(\Delta m\right)d\left(\Delta m\right),\label{3p4.5}
\end{equation}
\begin{equation}
\left\langle \Delta m\right\rangle _{-}=\int_{-\infty}^{0}\Delta mP\left(\Delta m\right)d\left(\Delta m\right).\label{3p4.6}
\end{equation}
We decompose the full distribution of magnification as a sum 
\begin{equation}
P\left(\Delta m\right)=\sum_{n=0}^{\infty}\mathcal{P}_{n}P_{n}\left(\Delta m\right),\label{3p4.7}
\end{equation}
where $\mathcal{P}_{n}$ is the probability of $n$ halo intersections
and $P_{n}\left(\Delta m\right)d\left(\Delta m\right)$ is the probability
of having a magnitude shift between $\Delta m$ and $\Delta m+d\left(\Delta m\right)$
given that there are $n$ intersections. 

The analytic expression for the magnitude shift for zero halo intersections
is \cite{p3Ref1} 
\begin{equation}
\Delta m=2.5\log_{10}\left|\left(1-\kappa_{{\rm interior}}\right)^{2}\right|,\label{3p4.8}
\end{equation}
 where 
\begin{equation}
\kappa_{{\rm interior}}\left(p\right)=-3\frac{H_{0}^{2}}{c^{2}}\Omega_{{\rm M}}f\left(z\right)\frac{y\left(y_{S}-y\right)}{y_{{\rm S}}a_{{\rm ex}}\left(z\right)}\sqrt{Y_{{\rm void}}^{2}-p^{2}}.\label{3p4.9}
\end{equation}
Here $f\left(z\right)$ is the fraction of the mass of the void on
the shell, $y$ is the comoving distance to the void, $y_{S}$ is
comoving distance to the source, $a_{ex}\left(z\right)$ is the scale
factor and $p$ is the comoving impact parameter. Note that for this
one void case, essentially all the negative $\Delta m$ contributions
are due to intersections with one halo and the positive $\Delta m$
contributions are due to the void interior (i.e., no halo intersections).
In this approximation, Eqs. (\ref{3p4.5}) and (\ref{3p4.6}) reduce to 
\begin{equation}
\left\langle \Delta m\right\rangle _{+}\simeq\mathcal{P}_{0}\int_{-\infty}^{\infty}\Delta mP_{0}\left(\Delta m\right)d\left(\Delta m\right)\label{3p4.10}
\end{equation}
 and
\begin{equation}
\left\langle \Delta m\right\rangle _{-}\simeq\mathcal{P}_{1}\int_{-\infty}^{\infty}\Delta mP_{1}\left(\Delta m\right)d\left(\Delta m\right).\label{3p4.11}
\end{equation}

We can compute $\mathcal{P}_{0}$ and $\mathcal{P}_{1}$ from Eq.
(\ref{3p4.3}) assuming $\mathcal{P}_{0}+\mathcal{P}_{1}=1$ for the one void
case, obtaining $\mathcal{P}_{1}=0.8$ and $\mathcal{P}_{0}=0.2$ which matches
with our simulations. The numerically computed means (4.5) and (4.6) of the magnified
and demagnified distributions agrees with their corresponding halo
and void interior theoretical values {[}computed from Eqs. (\ref{3p3.4}) -
(\ref{3p3.5}), (\ref{3p4.4}) \& (\ref{3p4.9}){]} to $\sim0.5\%$. We also numerically compute
the mean lensing convergence obtaining $-5\times10^{-4}$
magnitudes with standard deviation $2\times10^{-3}$
magnitudes. Thus the mean is consistent with zero as we would expect
from a general theorem.

\subsection{Shear}

So far in our analysis we have not included shear. We can include it as follows. The matrix $L_{\;C}^{A}$ for  the $j$-th void defined in Eq. (2.15c) of \cite{p3Ref1} is
\begin{equation}
L_{AB}=-2\int dy\left[\nabla_{A}\nabla_{B}\delta\Phi+{1\over2}\left(\delta\Phi\right)_{,yy}\delta_{AB}\right].\label{3p4.12}
\end{equation}
where $\delta \Phi$ is the potential perturbation, $y$ is comoving distance, the derivatives are with respect to comoving coordinates, and the integral is taken over just the $j$-th void. We decompose $L_{AC}$ into a trace part and a trace free part to obtain
\begin{equation}
L_{AB}=\frac{1}{w_{j}}\left[\kappa_{j}\delta_{AB}+\gamma_{jAB}\right],\label{3p4.13}
\end{equation}  
where $w_{j} = y_{j}\left(y_{S}-y_{j}\right)/y_{S}$, $y_{j}$ is the comoving distance to the $j$-th void, $\kappa_{j}$ is the lensing convergence we computed previously [Eqs. (\ref{3p3.3}) - (\ref{3p3.5}) and (\ref{3p4.9})], and the matrix $\boldsymbol{\gamma}_{j}$ is traceless. We compute the potential perturbation from the density perturbations (\ref{3p2.3}) and (\ref{3p2.4}), and insert into Eqs. (\ref{3p4.12}) and (\ref{3p4.13}) to obtain the shear term $\gamma_{jAB}$ which will be of the form $\gamma_{j}\epsilon_{AB}$ on a suitable choice of basis. For a single NFW halo the potential perturbation is 
\begin{equation}
\delta\Phi\left(r\right)=-4\pi G\rho_{0}R_{s}^{2}\left[\frac{R_{s}}{r}\ln\left(1+\frac{r}{R_{s}}\right)-\frac{1}{1+C}\right].\label{3p4.14}
\end{equation}

The shear due to the void interior and the halos on the shell is
\begin{equation}
\gamma_{\; B}^{A}=\left(\gamma_{{\rm void}}\right)_{\; B}^{A}+\sum_{i=1}^{N_{\rm{shell}}}\left(\gamma_{{\rm halo}}^i\right)_{\; B}^{A},\label{3p4.15}
\end{equation}
 where
\[
\left(\gamma_{{\rm void}}\right)_{\; B}^{A}=3\frac{H_{0}^{2}}{c^{2}}\Omega_{{\rm M}}f\left(z\right)\frac{y\left(y_{S}-y\right)}{y_{{\rm S}}a_{{\rm ex}}\left(z\right)}\frac{Y_{{\rm void}}}{\beta}\left(\delta_{B}^{A}-2\hat{p}^{A}\hat{p}_{B}\right)
\]
\begin{equation}
\left(\frac{2}{3}-\Theta\left(Y_{{\rm void}}-p\right)\left[\frac{2}{3}\left(1-\beta\right)^{3/2}+\beta\left(1-\beta\right)^{1/2}\right]\right),\label{3p4.16}
\end{equation}
and
\[
\left(\gamma_{{\rm halo}}^i\right)_{\; B}^{A}=\frac{\delta_{B}^{A}-2\hat{b}_{i}^{A}\hat{b}_{Bi}}{b_{i}^{2}}\Biggl[-4M_{{\rm halo}}
\]
\begin{equation}
\left.+16\pi\int_{b_{i}}^{CR_{s}}drr\rho_{{\rm halo}}\left(r\right)\sqrt{r^{2}-b_{i}^{2}}+8\pi\int_{b_{i}}^{CR_{s}}drr\frac{\rho_{{\rm halo}}\left(r\right)}{\sqrt{r^{2}-b_{i}^{2}}}\right].\label{3p4.17}
\end{equation}
Here $\mathbf{p} = p\hat{p}^{A}\mathbf{e_{A}}$ is again the comoving impact parameter to the void, $\beta=p^{2}/Y_{{\rm void}}^{2}$
and $\mathbf{b}_i = b_i\hat{b}_{i}^{A}\mathbf{e_{A}}$ is the physical impact parameter of the light ray to
the $i$-th halo. The first term in Eq. (\ref{3p4.17}) is the point mass contribution
of the halos. The second and third terms in Eq. (\ref{3p4.17}) are non zero only for
intersected halos. Evaluating the integrals using the NFW profile (\ref{3p2.1}), we find
that the intersected halo contribution $\left(\gamma_{{\rm int}}^i\right)_{\; B}^{A}$
is 
\begin{equation}
\left(\gamma_{{\rm int}}^i\right)_{\; B}^{A}=\frac{16\pi G}{c^{2}}a_{{\rm ex}}\left(z\right)\rho_{0}R_{s}\frac{y\left(y_{S}-y\right)}{y_{S}}\left(\delta_{B}^{A}-2\hat{b}_{i}^{A}\hat{b}_{Bi}\right)\gamma_{h}.\label{3p4.18}
\end{equation}
Here for $\alpha<1$
\[
\gamma_{h}=\frac{1}{\alpha}\left\{ \frac{\left(2-\alpha\right)\sqrt{C^{2}-\alpha}}{2\left(\alpha-1\right)\left(C+1\right)}+\ln\left[\frac{C+\sqrt{C^{2}-\alpha}}{\sqrt{\alpha}}\right]\right.
\]
\[
+\frac{\left(3\alpha-2\right)}{\left(1-\alpha\right)^{3/2}}\left(\tanh^{-1}\left[\sqrt{\frac{1-\sqrt{\alpha}}{\sqrt{\alpha}+1}}\right]\right.
\]
\begin{equation}
\left.\left.-\tanh^{-1}\left[\frac{\sqrt{1-\alpha}}{C+1+\sqrt{C^{2}-\alpha}}\right]\right)\right\} ,\label{3p4.19}
\end{equation}
and for $ $$1<\alpha<C^{2}$
\[
\gamma_{h}=\frac{1}{\alpha}\left\{ \frac{\left(2-\alpha\right)\sqrt{C^{2}-\alpha}}{2\left(\alpha-1\right)\left(C+1\right)}+\ln\left[\frac{C+\sqrt{C^{2}-\alpha}}{\sqrt{\alpha}}\right]\right.
\]
\[
-\frac{\left(3\alpha-2\right)}{\left(\alpha-1\right)^{3/2}}\left(\tan^{-1}\left[\sqrt{\frac{\sqrt{\alpha}-1}{\sqrt{\alpha}+1}}\right]\right.
\]
\begin{equation}
\left.\left.-\tan^{-1}\left[\frac{\sqrt{\alpha-1}}{C+1+\sqrt{C^{2}-\alpha}}\right]\right)\right\} ,\label{3p4.20}
\end{equation}
where $\alpha=b_i^{2}/R_{s}^{2}$.

From Eqs. (\ref{3p4.18}) - (\ref{3p4.20}),
$\gamma^{2}\sim1/b^{4}$. Therefore contributions from shear are heavily
suppressed. In our numerical analysis, the standard deviation changes by less than $3\%$ 
if shear is neglected.

\subsection{Qualitative features of magnification distributions}

With the accuracy of our method tested, we now explore the magnification
distributions in more general situations with many voids, distributed
along the line of sight with random impact parameters according to
the algorithm discussed in \cite{p3Ref1}. For example, for sources at redshift
$z_{{\rm s}}=1$, there are 47 voids of comoving radius $Y_{{\rm void}}=35$
Mpc along the line of sight. We follow steps 1 to 8 of Section IIC
of our previous paper \cite{p3Ref1}, but with the modification that the
matrices $\mathbf{J}$, $\mathbf{K}$, $\mathbf{L}$ and $\mathbf{M}$ now incorporate the effects of the halo substructure
of the shell.

In Figure 4, we plot the log of the magnification distribution for
$z_{s}=0.5,\:1.0\:$ and $1.5$. In our SRN model, we have voids with randomly distributed halos on their surface
and a smooth interior. We denote by $\mathcal{P}_{n}$ the probability
of having $n$ halo intersections. The whole probability distribution
can be decomposed into a sum of probability distribution for different
numbers of halo intersections, like in Eq. (\ref{3p4.7}). We
plot the 25\%, 50\% and 75\% quartiles of the distributions as horizontal
lines (top, middle and bottom respectively). For high redshifts, most
of the probability is concentrated in the demagnified areas where
the rays hit only a few halos or simply pass through without hitting
any.

\begin{figure}
\includegraphics[scale=0.54]{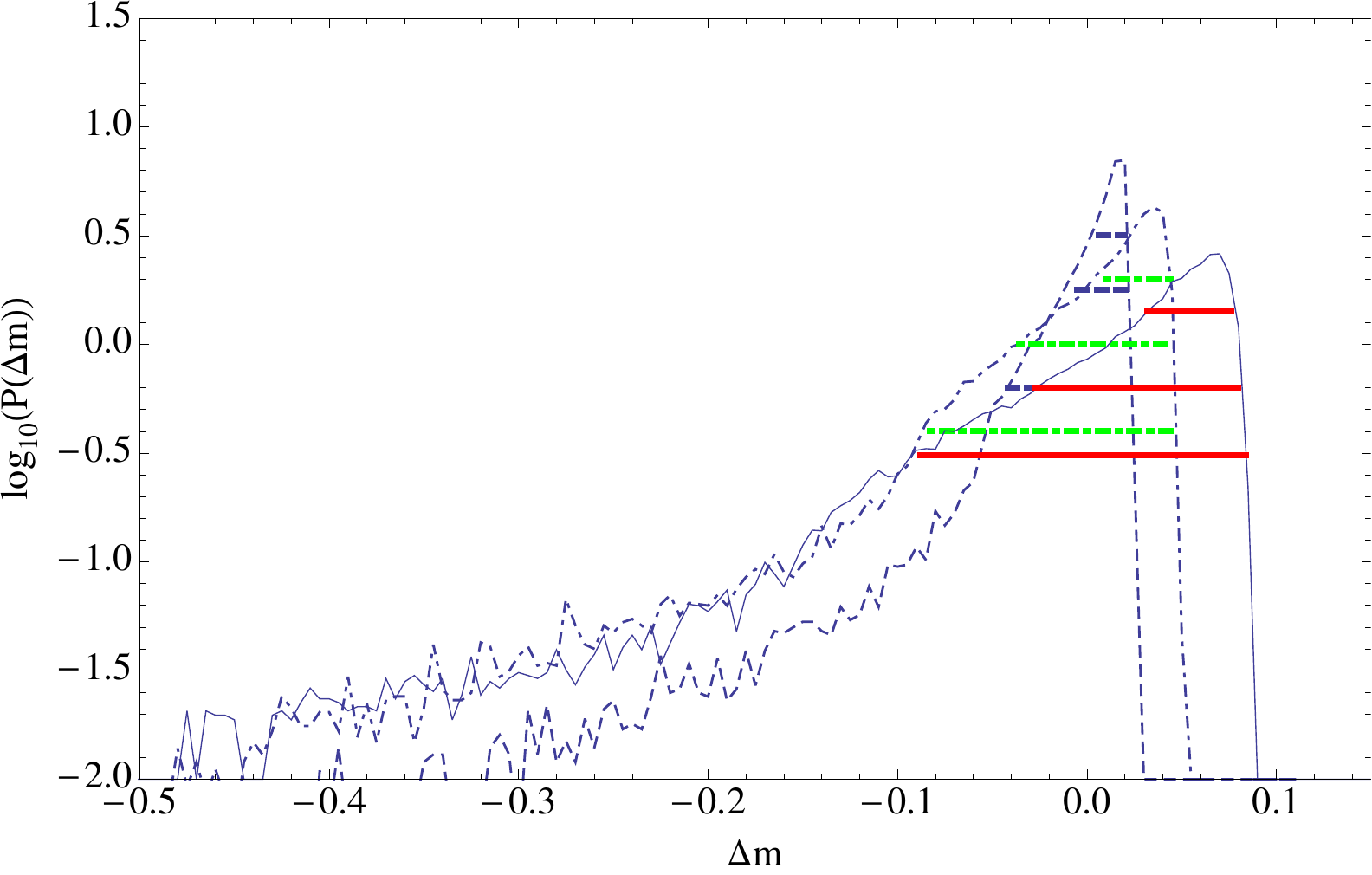}

\caption{The probability distributions of magnitude shifts $\Delta m$ for
the SRN model with sources at redshifts of $z_{s}=0.5$ (dashed),
$z_{s}=1.0$ (dot-dashed) and $z_{s}=1.5$ (solid), for comoving voids
of radius R = 35 Mpc with 90\% of the void mass on the shell today.
The horizontal lines are the 25\% (top), 50\% (middle) and 75\% (bottom)
quartiles about the peak of the distribution.}
\end{figure}

Note that the total probability in the tail on the magnification side
$\left(\Delta m<0\right)$ increases with redshift, because of the
increased probability of hitting halos at higher redshifts. For example,
at $\Delta m=-0.2$, we would expect the probability density for $z_{{\rm s}}=1.5$
to be roughly 2-3 times as large as the probability density for $z_{{\rm s}}=0.5$
because the number of voids that rays have to pass through in the
former case is 62 where as for the latter it is 27. In addition, rays
at high redshifts have more close encounters with halos that generate
shear.

Figure 5 shows the standard deviation of the distribution, $\sigma_{m}$,
as a function of redshift of the source, $z_{{\rm s}}$. This standard deviation for voids and halos is $\sim3$ times
larger at $z_{{\rm s}}=1$ than that for a model with mass compensated voids with
a shell thickness of 1 Mpc and no halos [1]. We note that most of the
contribution to the standard deviation come from rays that intersect halos.
Also, the standard deviation we compute
agrees well with that computed using other methods. For example, our standard deviation
for $z_{{\rm s}}=1.5$ is $\sigma_{m}=0.072$, which agrees to within
20\% with the standard deviation of the distribution shown in Figure
1 of Ref. \cite{p3Ref5}. We compare our results to those obtained using another method introduced
in Refs. \cite{p3Ref11, p3Ref12} in the next subsection.

In Appendix B we derive the following approximate result for the standard deviation: 
\[
\sigma_m={5\over{\log\left(10\right)}}\left({\Omega_M\over 2}\gbarh\sum_j
{(H_0Y_{\rm{void}}f_j)H_0^2w_j^2}\right.
\]
\[
\;\;\;\;\;\;\;\;\;\;\;\;\;\;\;\;+\left.{\Omega_M^2\over 2}\sum_j{(H_0Y_{\rm{void}}f_j)^2H_0^2w_j^2\over a_j^2}\right)^{1\over2}.
\tag{{4.21}}
\label{sigm}
\]
Here $\gbarh$ is a dimensionless parameter which represents the contribution from halos whose detailed form is given by Eq. (B21), $Y_{\rm{void}}$ is the comoving radius of the voids, $f_j=f\left(z_j\right)$ is the fraction of the mass of $j$-th void on its surface and $a_j=a\left(z_j\right)$ is the scale factor. The result (4.21) assumes statistical independence of halos within voids and also of voids from one another and neglects lensing shear. There are two main qualitative features of the result (4.21). First, the contribution to the standard deviation due to the halos [the first term in Eq. (4.21)] depends primarily on their gravitational potential, and the contribution due to voids [the second term in Eq. (4.21)] depends primarily on the size of the underdense core. Second, the halo contribution is bigger than the interior contribution and hence the standard deviation is dominated by halos. For example, using the above expression, the ratio of the contribution due to the halos to the contribution due to the core is $\sim 100$ for $z_s=1$ and for the void and halo parameters defined in Table II.

We discuss further the analytic calculation of standard deviation without shear in Appendix B. Our numerical results agree with these approximate analytic predictions to within $\sim20\%$.

\begin{figure}
\includegraphics[scale=0.58]{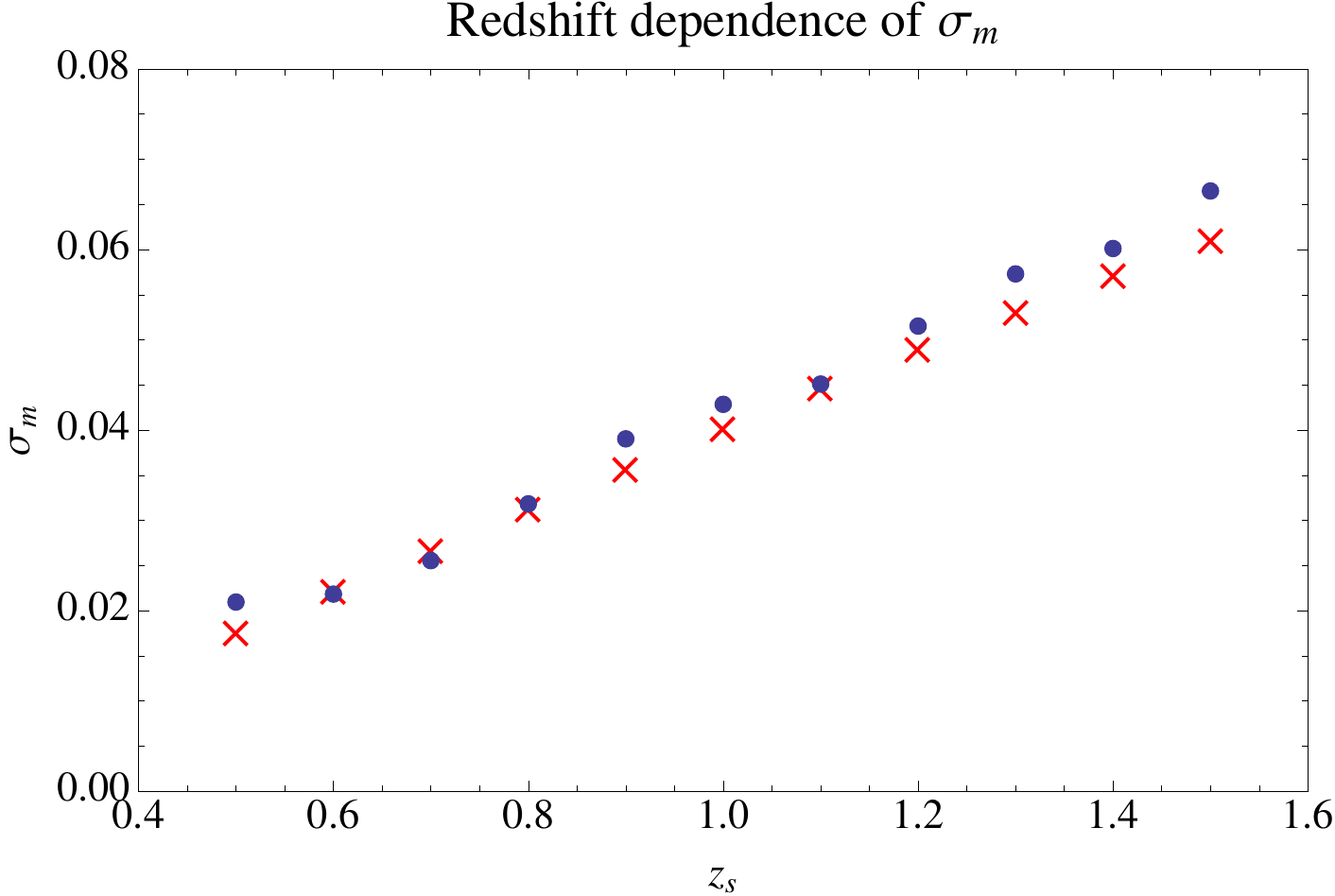}

\caption{Redshift dependence of standard deviation of distribution of magnitude
shifts, for comoving voids of radius R = 35 Mpc with 90\% of the void
mass on the shell today. The crosses are analytic results.}
\end{figure}

\begin{figure}
{\includegraphics[scale=0.53]{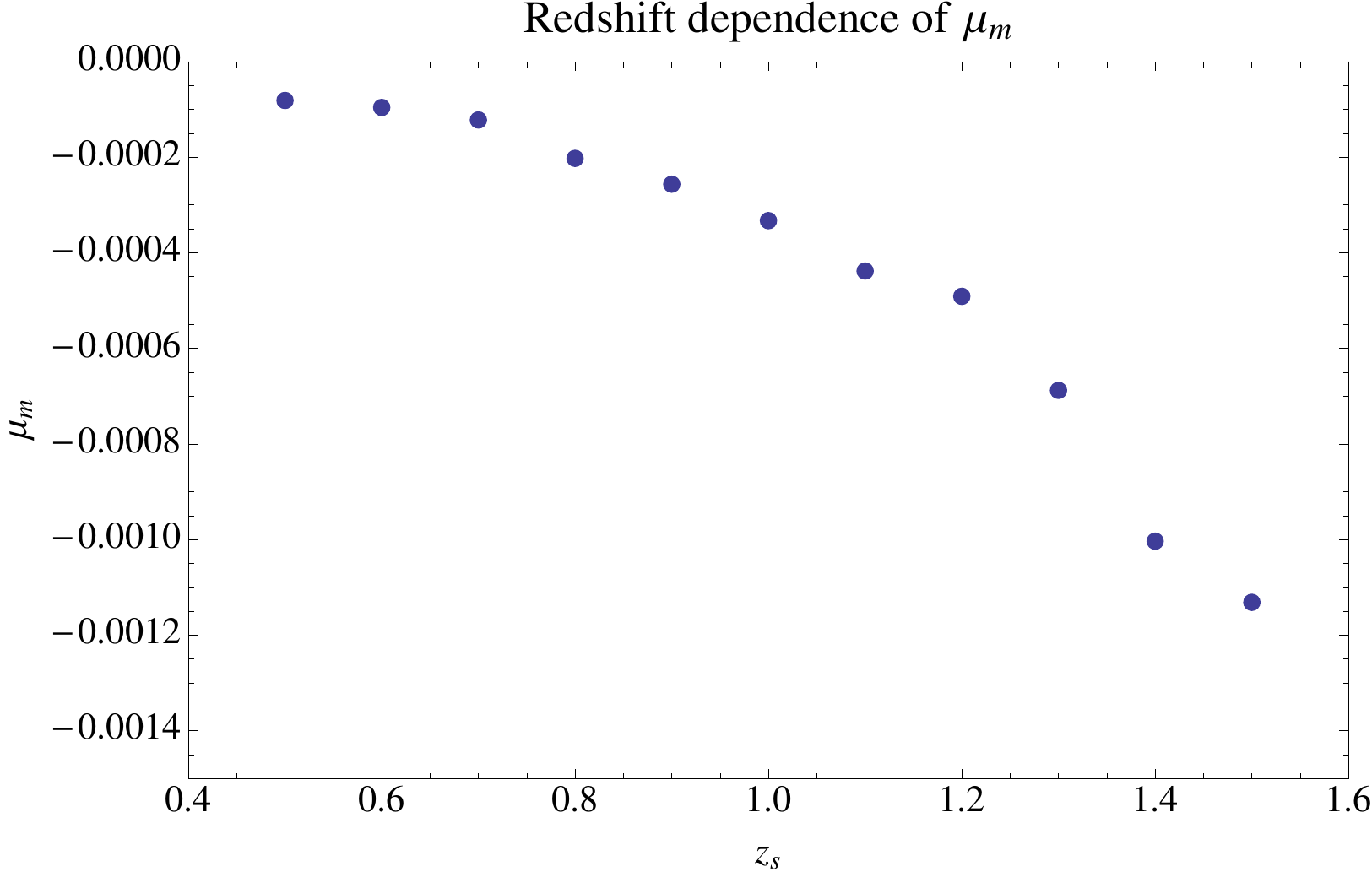}

}

{\includegraphics[scale=0.60]{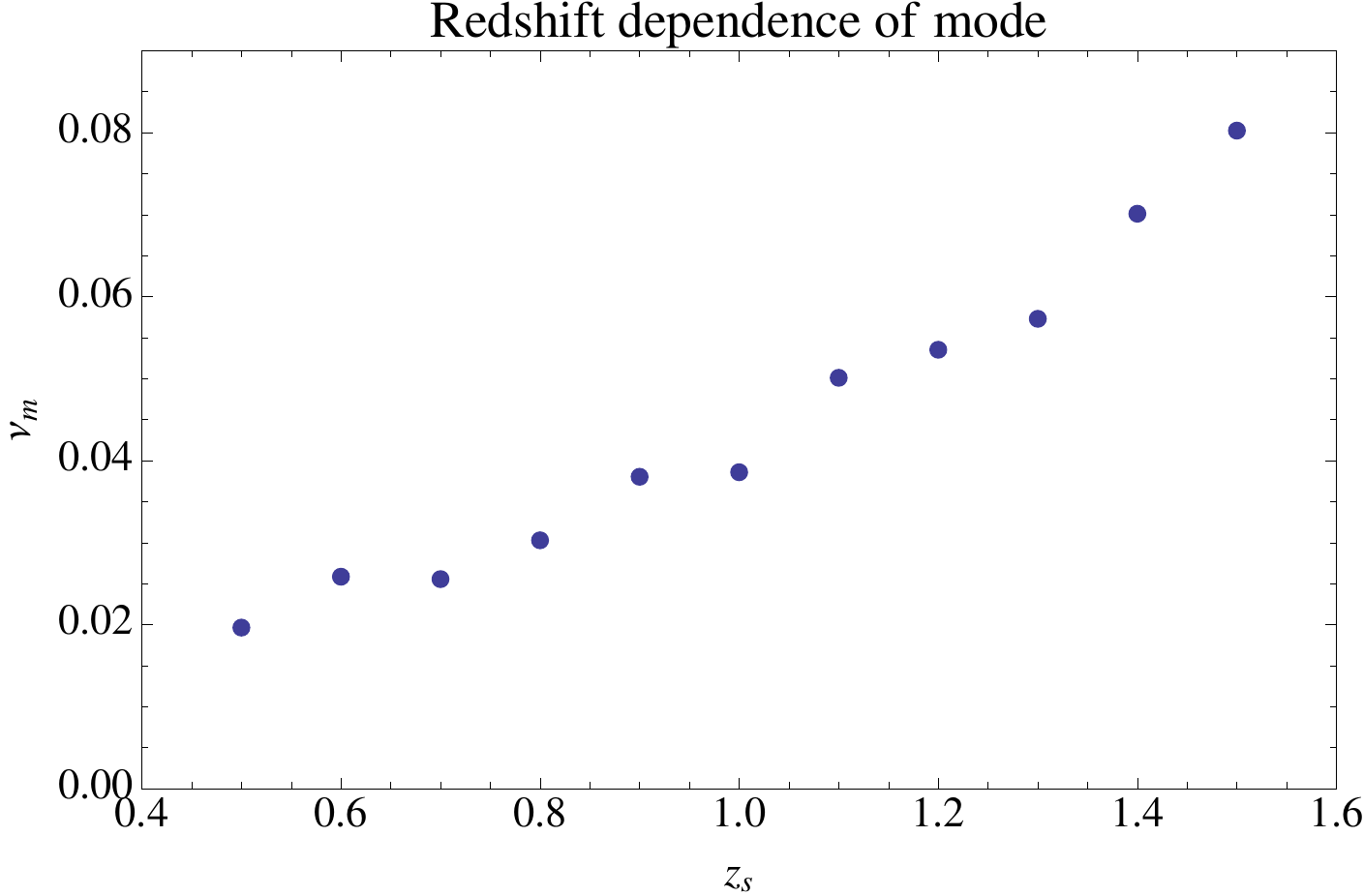}

}

\caption{Plot of mean (top) and mode (bottom) of the distribution as a function of redshift
in SRN. The mode takes on increasingly positive values with redshift and the mean is increasingly negative with redshift.}
\end{figure}

\subsection{Redshift dependence of mean and mode of magnitude shift}

While the mean of lensing convergence
vanishes, the mean magnitude shift does not, because magnitude shift
is a nonlinear function of $\kappa$, defined in Eq. (\ref{3p4.4}). Figure
6a shows the mean $\mu_{m}$ of the distribution of magnification
shifts $\Delta m$, which increases with redshift as $\propto \sigma_m^{2}$.
This is the expected theoretical behavior: for small values of
$\kappa$ and ignoring shear, we can approximate Eq. (\ref{3p4.4}) as 
\begin{equation}
\Delta m\simeq\frac{5}{\ln10}\ln\left|\left(1-\kappa\right)\right|\simeq\frac{5}{\ln10}\left(-\kappa-\frac{1}{2}\kappa^{2}\right).\label{3p4.21}
\end{equation}
The mean magnitude shift is then proportional to the mean of the square
of $\kappa$ as the mean of $\kappa$ is vanishing, 
\begin{equation}
\mu_{m}\simeq-\frac{2.5}{\ln10}\left\langle \kappa^{2}\right\rangle .\label{3p4.22}
\end{equation}
The standard deviation, from Eq. (\ref{3p3.5}) in \cite{p3Ref1} simplifies to 
\begin{equation}
\sigma_{m}=\frac{5}{\ln10}\sqrt{\left\langle \kappa^{2}\right\rangle },\label{3p4.23}
\end{equation}
and so $\mu_{m}\simeq -0.23\sigma_{m}^{2}$ which agrees with our numerical
results to within $\sim 10 \%$.

On average there is a small overall magnification of light beams. Figure 6b shows the mode $\nu_{m}$, the location of the maximum of the PDF, which also increases
with redshift. The modes of the magnification to redshift 1.5 are positive because an overall demagnification
occurs for most of the light rays as they pass
through the interior of the voids while hitting halos. Note that
the modes are larger than the corresponding means.

\begin{figure}
{\includegraphics[scale=0.53]{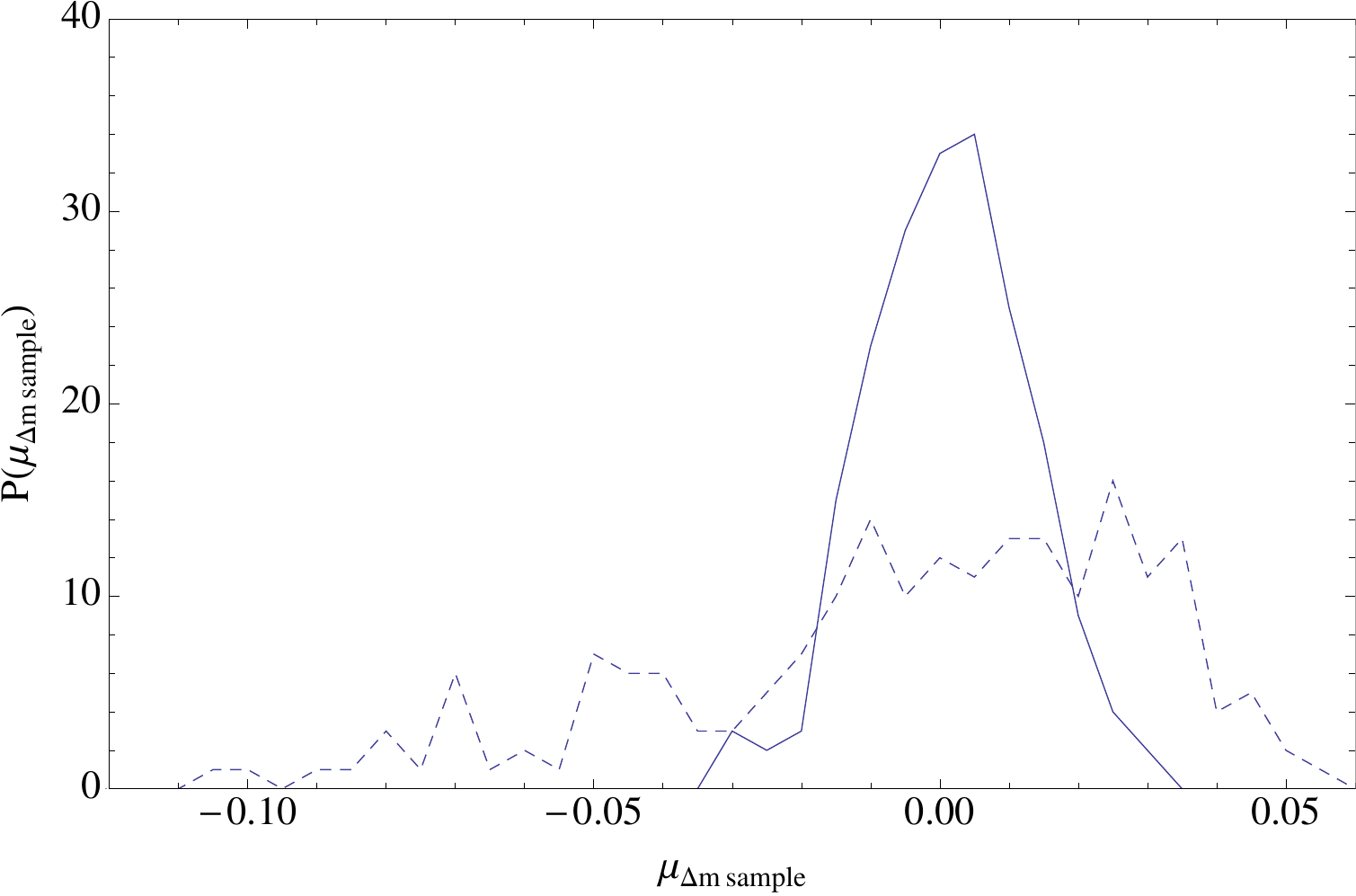}}

{\includegraphics[scale=0.53]{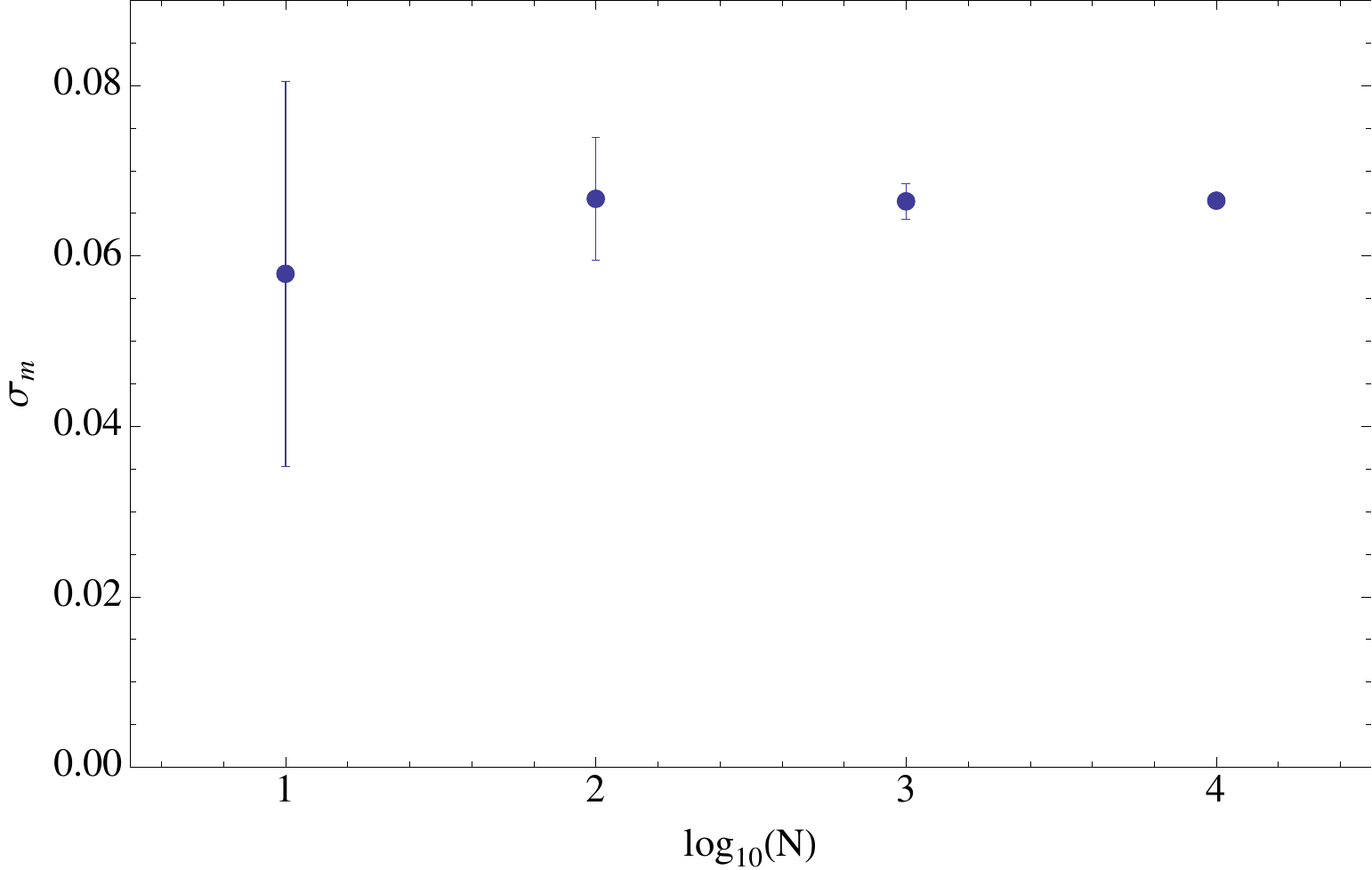}}

\caption{Top: Plot of distribution of means of 200 samples of 10 (dashed) and 100 (solid) sources each at
redshift $z_{s}=1.5$. The mean of the distribution of means is 0.011
magnitudes and the standard deviation is 0.028 magnitudes for the sample of 10. The respective numbers for the sample of 100 sources is 0.0004 and 0.011, showing that more sources reduces demagnification bias. Bottom: Plot of the range of standard deviation for samples of different sizes showing convergence as $N\to10^4$.}
\end{figure}

In realistic surveys, one can expect to find only a few standard candle
sources for every redshift or every redshift bin. This severely constrains
the accuracy of the cosmological parameters we can infer from such
observations. To illustrate the extent of lensing degradation in measuring
cosmological parameters, we pick 200 sets of randomly placed 10 or 100 sources
at $z_{s}=1.5$. We find the mean of each of these samples in the
set and plot the resulting distribution of means in Figure 7 (top). The mean of the means for the 10 sources case is  0.011 magnitudes and the standard deviation of the means is 0.028 magnitudes. The respective numbers for the 100 sources case are 0.0004 and 0.011 magnitudes. A change in cosmological parameters by $1 \%$ implies a change in $\Delta m$ of 0.015 magnitudes. Thus for data acquired from surveys, the lensing
degradation is quite a significant effect, although it can be mitigated
by increasing the number of sources. This is also seen in Figure 7 (bottom) where we plot the range of standard deviation for samples of different sizes. For a large enough sample, the bias in magnification can be accurately taken into account. This effect is studied in \cite{p3Ref5}
which shows that lensing degradation effectively decreases the number
of useful supernovae by a factor of 3 at source redshift 1.5.

Our work is broadly consistent with other work, \cite{p3Ref11, p3Ref12, p3Ref14, p3Ref15, p3Ref16, p3Ref17, p3Ref18, p3Ref19, p3Ref20, p3Ref21, p3Ref22} in
this area. A similar computational method has been developed by Kainulainen
\& Marra, Refs. \cite{p3Ref11, p3Ref12}. Their model consists of filaments and
halos of various sizes, where the mass fraction in filaments is 0.5
and the rest is distributed in halos. To compare with their results,
we use the SRN model and choose parameters to match their cosmology,
i.e., $\Omega_{{\rm M}}=0.25$, $z_{s}=1.5$, $H_{0}=73$ $\mathrm{kms^{-1}Mpc^{-1}}$
and $f=0.5$. We do not include shear for this comparison as it is
neglected in their analysis. Our magnification PDF qualitatively agrees
with that of Kainulainen \& Marra as shown in Figure 8.

\begin{figure}
\includegraphics[scale=0.53]{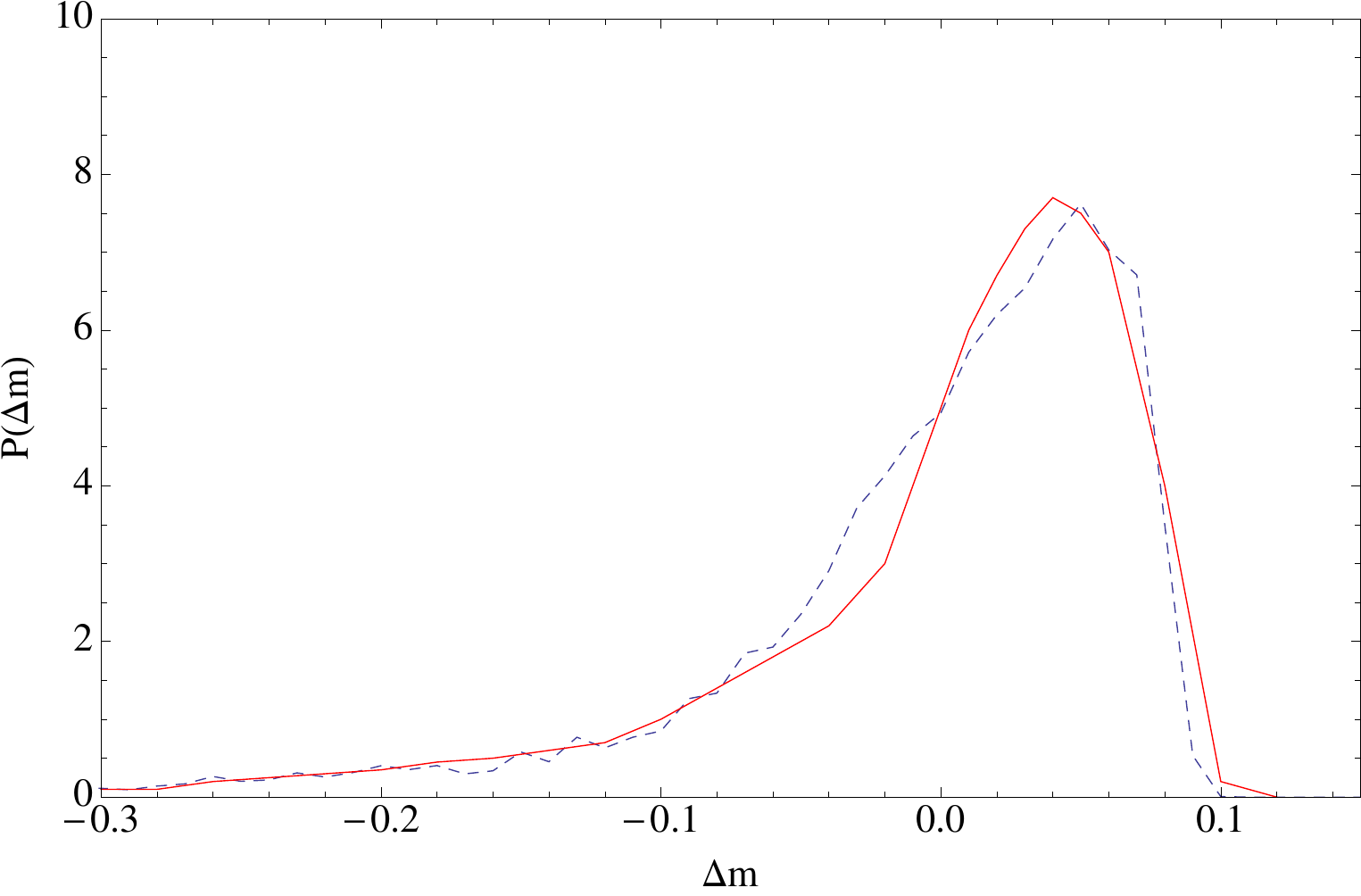}

\caption{The probability distributions of magnitude shifts $\Delta m$ for
an SRN model (dashed lines) with sources at redshift $z_{s}=1.5$,
comoving voids of radius 35 Mpc, with 50\% of the void mass in halos
today, $\Omega_{{\rm M}}=0.25$, $\Omega_{\Lambda}=0.75$, $H_{0}=73$
$\mathrm{kms^{-1}Mpc^{-1}}$ and with no shear, compared to the results
in Figure 5 (we reproduced the plot by picking points from their figure)
of the model in Kainulainen and Marra \cite{p3Ref12} where they have 50\%
of mass in halos and all other parameters same as ours. The two distributions
are qualitatively similar.}
\end{figure}

\section{Results for Swiss Raisin Raisin model}

In our Swiss Raisin Raisin (SRR) model, in addition to replacing the
smooth surface density on the shell with a collection of halos, the
mass in the interior is also broken up into NFW halos with the same
parameters as before. The parameters of the SRR model are listed in
Table III.

\begin{table}
\begin{tabular}{|c|c|}
\hline 
\multicolumn{1}{|c||}{Quantity} & Value\tabularnewline
\hline 
\hline 
$\Omega_{{\rm M}}$ & 0.3\tabularnewline
\hline 
$\Omega_{\Lambda}$ & 0.7\tabularnewline
\hline 
$H_{0}$ & 70 $\mathrm{kms^{-1}Mpc^{-1}}$\tabularnewline
\hline 
$Y_{{\rm void}}$ & $35$ Mpc\tabularnewline
\hline 
Halo profile & NFW\tabularnewline
\hline 
Present fraction of void mass on shell & 0.9\tabularnewline
\hline 
Fraction of shell mass in halos & 1.0\tabularnewline
\hline 
Fraction of interior mass in halos & 1.0\tabularnewline
\hline 
\end{tabular}

\caption{Parameters of SRR model}

\end{table}

\subsection{A single void }

In this section we focus on a single void with no shear. Again, we
use Eqs. (\ref{3p2.4}), (\ref{3p3.4}) - (\ref{3p3.5}) and (\ref{3p4.8}) to compute the magnifications.
For a 35 Mpc void and using the same halo parameters as in Section
II, the intersection probability remains about the same as in the
previous model. The change due to the addition of a few halos in the
vast interior region is negligible. The expected number of halo intersections
$N_{\rm int}\left(z\right)$ is given by Eq. (\ref{3p4.3})
with the shell mass replaced by the total void mass $M_{\rm void}$
\begin{equation}
N_{\rm int}\left(z\right)=\frac{M_{\rm void}R_{{\rm halo}}^{2}}{Y_{{\rm void}}^{2}a_{{\rm ex}}^{2}\left(z\right)M_{{\rm halo}}}\label{3p5.1}
\end{equation}
Using the same parameters as in the one void case
in the SRN model we obtain $N_{\rm int}\left(0.45\right)\simeq0.88$. We find that light
rays hit halos 8720 times for $N = 10^4$ runs which translates to an intersection probability
of $\sim88\%$ and agrees well with Eq. (\ref{3p5.1}).

Again we find a bimodal distribution of magnitude shifts. In Figure
9, this bimodal distribution is superimposed on the SRN plots from
Figure 3. Since the density contrast in the interior is increased
by $\sim10\%$ (if halos are not hit) compared to our SRN model, the
demagnified distribution shifts towards the right. Figure 9a shows
this shift in underdense part of the distribution. Figure 9b is the
$\Delta m<0$ part of the distribution. This accounts for roughly
88\% of the total distribution and it is similar to the distribution
in Figure 3b.

\begin{figure}
{\includegraphics[scale=0.52]{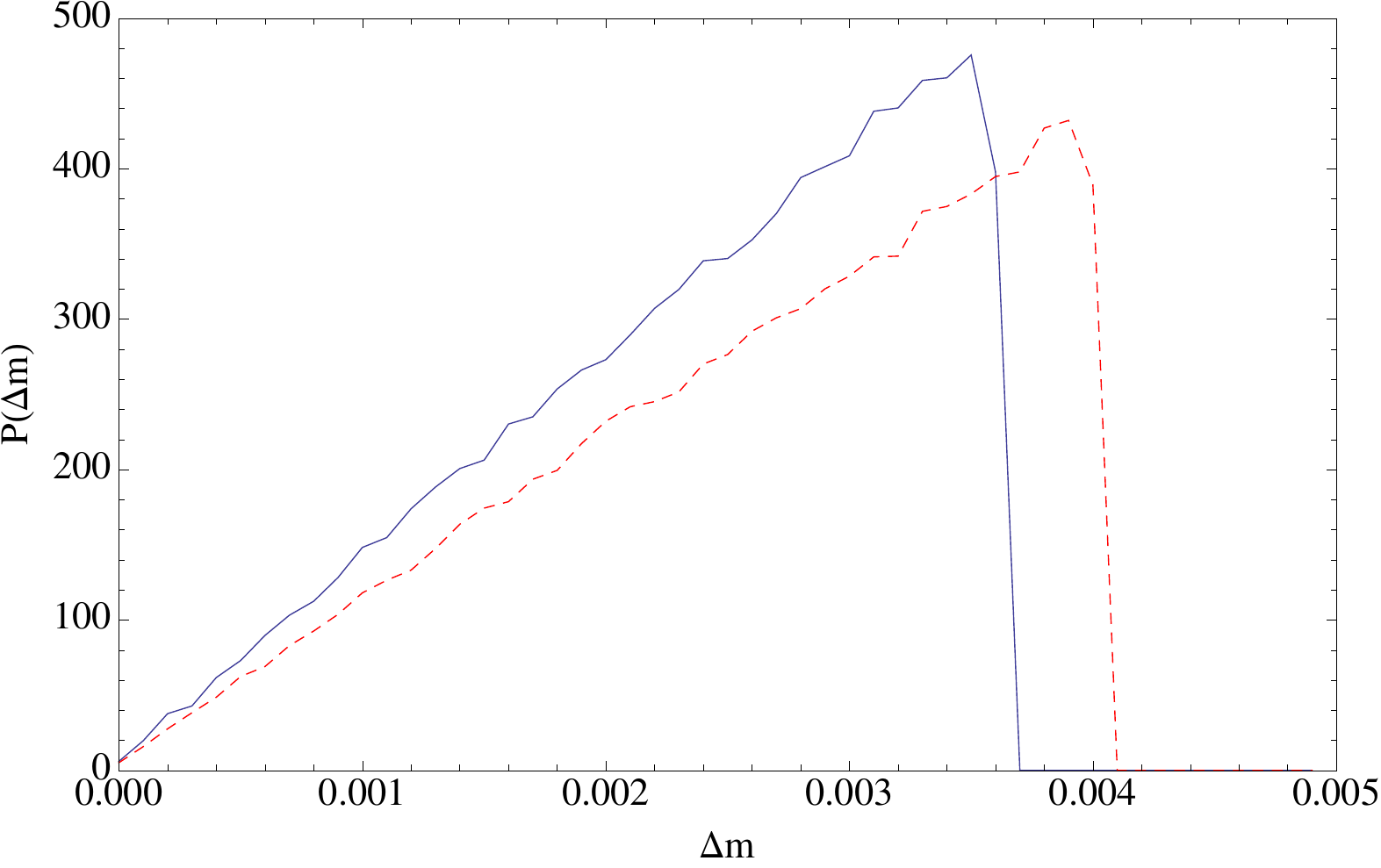}

}

{\includegraphics[scale=0.52]{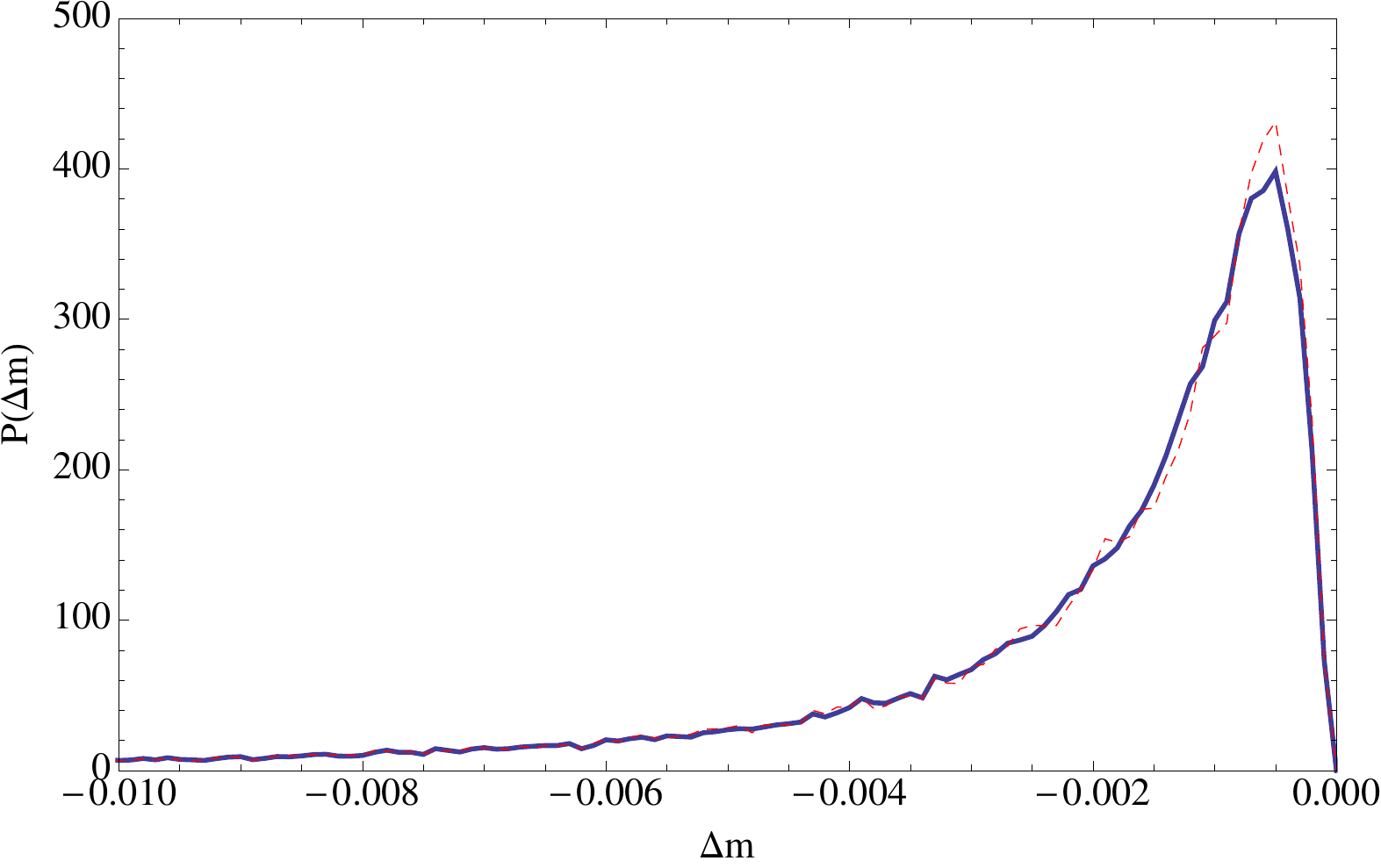}

}

\caption{The probability distribution of magnitude shifts $\Delta m$ for a
simulation in the SRR model (dashed line) superimposed on the corresponding probability distribution in the SRN model (solid line) with $\Omega_{M}=0.3$, with one void
of radius 35 Mpc and fraction of void mass on the shell today $f=0.9$.
Note that the demagnified part, (top), is shifted because there is an
increase in the density contrast in the interior of voids, while the
magnified part, (bottom), is mostly unchanged.}
\end{figure}

\begin{figure}
\includegraphics[scale=0.6]{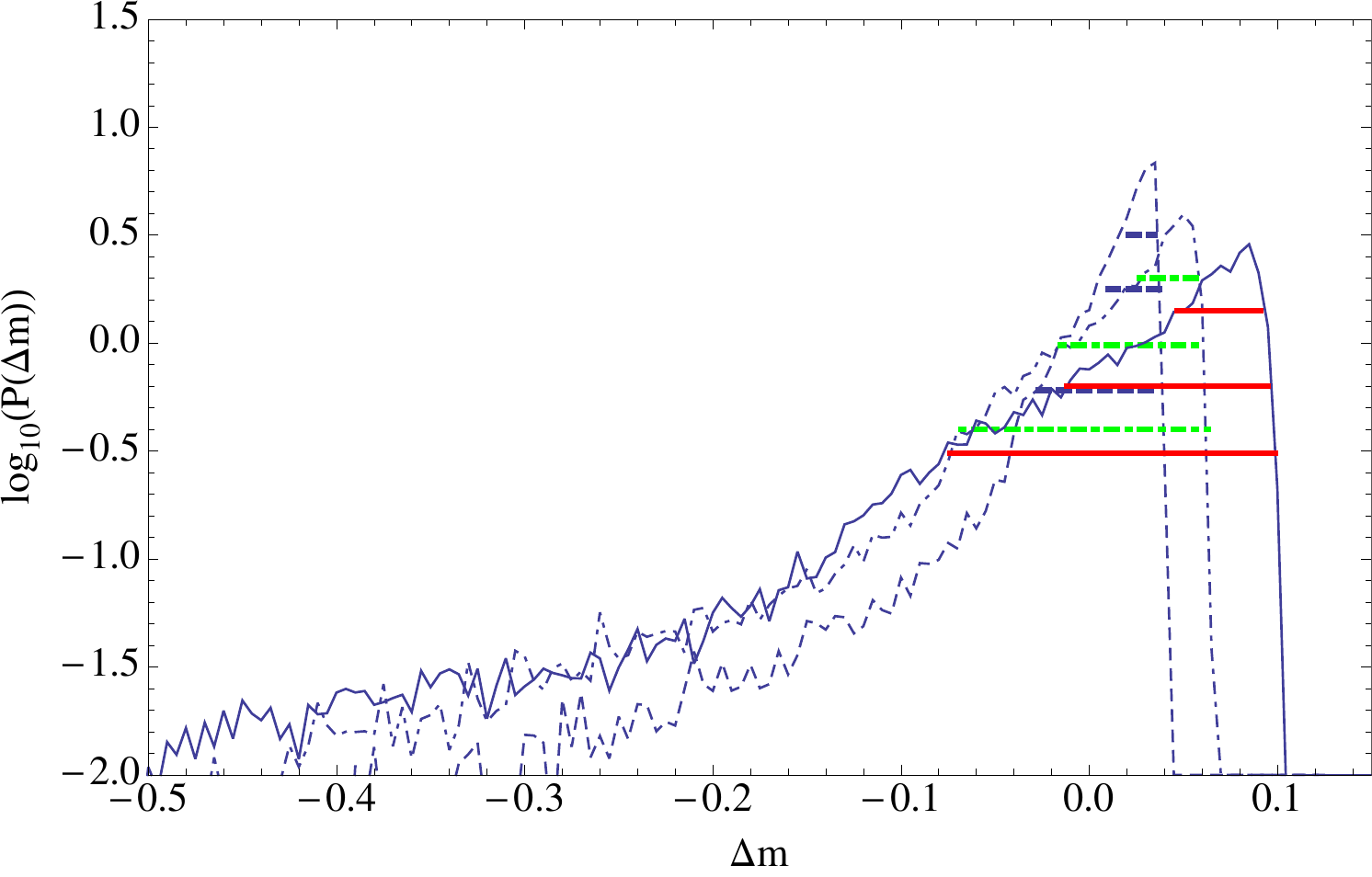}

\caption{The probability distributions of magnitude shifts $\Delta m$ for
the SRR model with sources at redshifts of $z=0.5$ (dashed), $z=1.0$
(dot dashed) and $z=1.5$ (solid), for voids of comoving radius R
= 35 Mpc, with 90\% of the void mass on the shell today. We see the
same qualitative features as in SRN for the corresponding redshift
but all the distributions are shifted towards demagnification. The
horizontal lines are the 25\% (top), 50\% (middle) and 75\% (bottom)
quartiles of the distribution from the peaks.}
\end{figure}

\subsection{Redshift dependence of distributions}

Next we explore the bias (i.e., the mean of the distribution) due
to halos and voids for sources at various redshifts, namely, for $z_{{\rm s}}=0.5,\:1.0\:$
and $1.5$. To compute the shear in this case, for the void contribution
we use Eq. (\ref{3p4.9}) but with $f\left(z\right)=1$, and in Eqs.
(\ref{3p4.18}) - (\ref{3p4.20}), we sum over both the surface and interior halos.
The magnifications shown in Figure 10 are predominantly due to halos
while the mostly empty interior has a demagnifying effect. Due to the increased underdensity inside the void,
the magnitudes shift towards demagnification. The standard deviation
for $z_{s}=1.5$ is $\sigma_{m}=0.072$. Again we note that the modes
are larger than the means and also shift towards the demagnification
end of the plot with increasing redshift. The tails of these distributions
are similar to the ones obtained in the SRN model in Figure 4.

Next we consider the mean magnitude shift, $\mu_{m}$, and its mode,
$\nu_{m}$, in the two models. The key feature here is that the
underdense interior is more prominent in the SRR model. We expect
that the mean magnitude shift and its mode should shift and the difference
in the means should increase with redshift. In Figure 11, we plot
the means and modes for the two models and we show that $\mu_{m}$
for SRR is $\sim10 - 20\%$ greater than that for SRN at $z_{{\rm s}}=1$. 

\begin{figure}
{\includegraphics[scale=0.56]{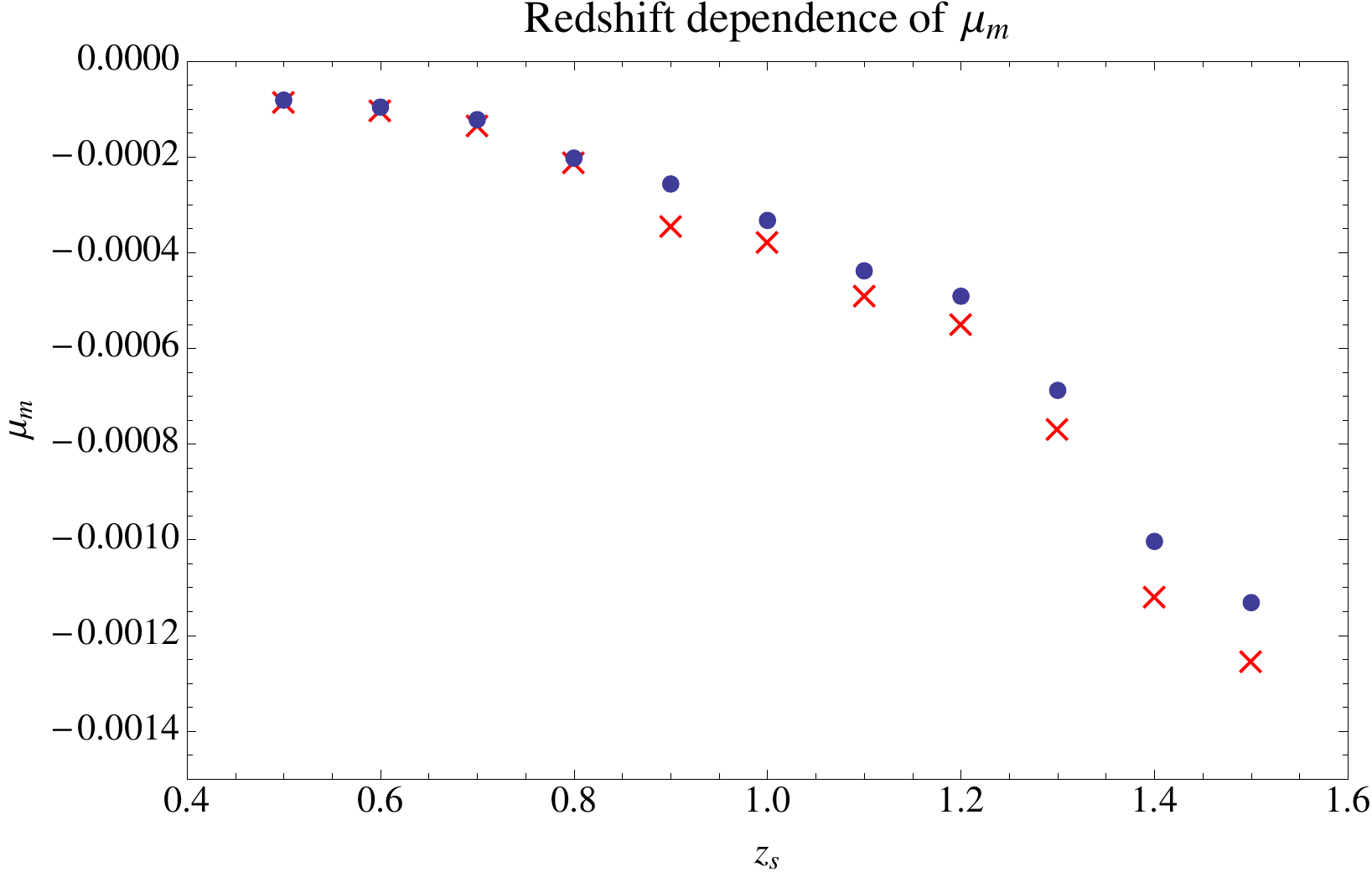}

}

{\includegraphics[scale=0.61]{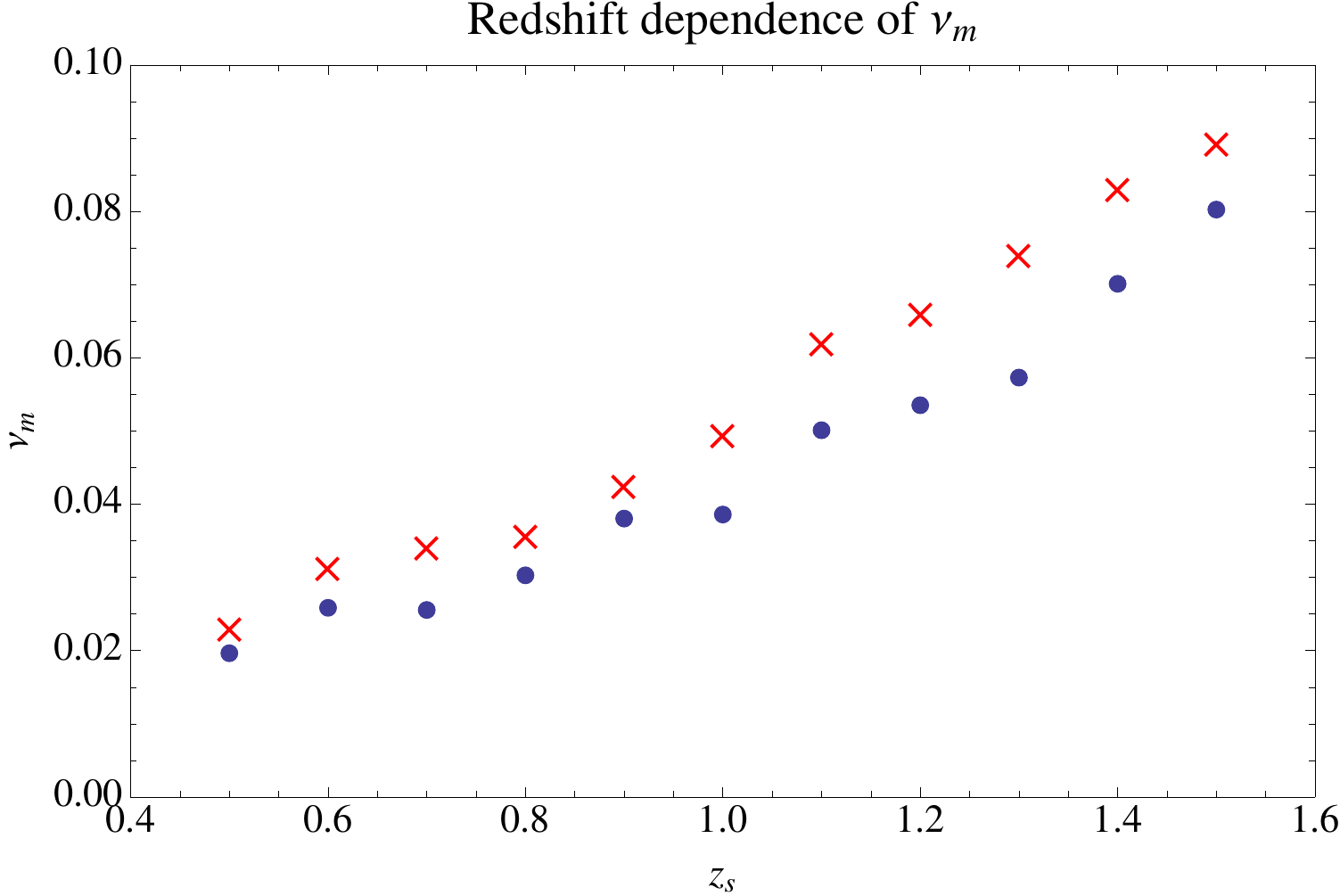}

}

\caption{The mean and mode of magnification shift for the two models - Points:
SRN; Crossed points: SRR. Top: The mean of magnification shift, $\mu_{m}$, for the two models. We
see that the difference in the means increases with redshift and at
$z=1$ and it is $\sim10\%$. Bottom: The mode of magnification shift, $\nu_{m}$, for the two models. We
see that the difference in the means increases with redshift and at
$z=1$ and it is $\sim20\%$.}
\end{figure}

These two models are interesting because they represent the two possible extremes
of the matter distribution in voids - one where the matter is smoothly distributed with no structure and another with only chunky
NFW halos. In reality, the underdense region will be composed of both
halos and an ambient intergalactic medium. By studying the completely
smooth interior case (SRN) and the completely granular interior case
(SRR), we expect to bracket the true distribution.

\subsection{Effect of large scale structure}

\begin{figure}
\includegraphics[scale=0.59]{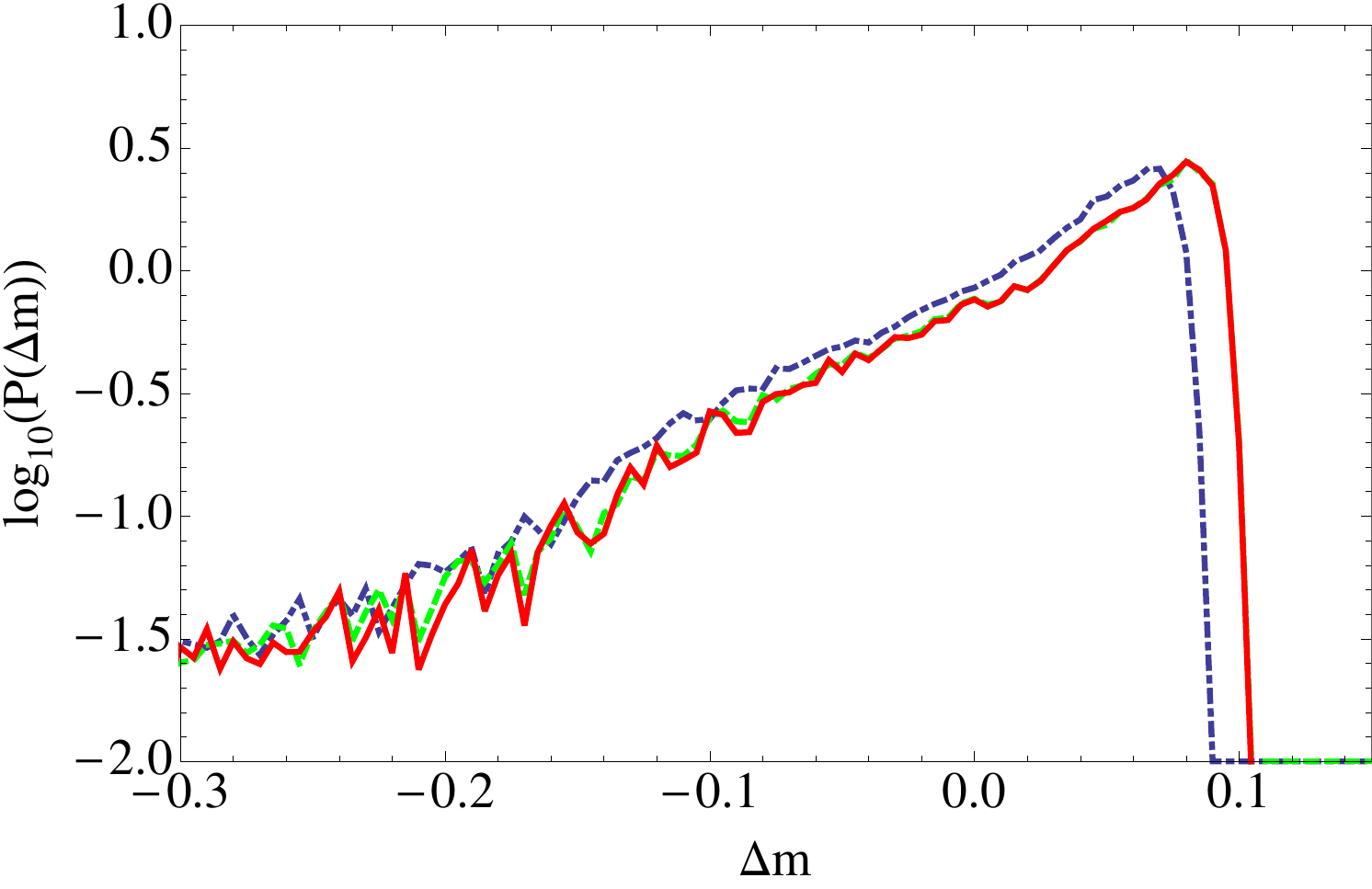}

\caption{The probability distributions of magnitude shifts $\Delta m$ for
simulations in SRR (solid), the SRN model (dot dashed), for comoving
voids of radius R = 35 Mpc with $f=0.9$ today. Also shown is the
model with uniformly distributed halos everywhere and with no voids
(dashed - this is the $f=0$ limit of the SRR model) with source at
redshifts of $z=1.5$. The distributions are similar in the SRR models
because the expected number of halos intersected is independent of
of the radial distribution of halos, where as the distribution in
the SRN model is qualitatively similar but is shifted towards negative
magnifications. }
\end{figure}

Previous studies in the literature have modeled the magnification
effects of only voids \cite{p3Ref1, p3Ref7, p3Ref10} or only halos \cite{p3Ref3, p3Ref5, p3Ref11},
while other studies have considered very specific models with a particular
distribution of filaments and halos \cite{p3Ref12}. In our work, we present
two models that incorporate the effects of both halos and voids. We have considered cosmological models where at large
scales matter evolves to cluster on the edges of spherical voids.

One limit of our model is the case where there are no halos on the surface and the interior is composed entirely of halos.
This corresponds to the $f=0$ limit of the SRR model, a "no void limit". In Figure 12,
we compare the magnification distribution we obtain for this configuration
for $z_{s}=1.5$ to the corresponding distribution in
the SRR and SRN models, for comoving voids of radius R = 35 Mpc with
90\% of the void mass on the shell today. From Eq. (\ref{3p5.1}), the expected number of halo intersections in the SRR
model is independent of the radial distribution of the halos and hence
the SRR and no void distributions should be similar, as observed.
However, in the SRN model the number of halo intersections is lower
due to the smaller number of halos. Hence the distribution is shifted
towards negative magnification shifts.

\section{Conclusions}

In this paper, we presented two simple models to study the effects
of both voids and halos on distance modulus shifts due to gravitational
lensing. Our results may be useful for future surveys that gather
data on luminosity distances to various astronomical sources to constrain
properties of the source of cosmic acceleration. The core of our model
is constructed by considering a $\Lambda$CDM Swiss cheese cosmology
with mass compensated, randomly located voids, while our small scale
halo structures are non-evolving and chosen to be all the same size
and with an NFW matter profile. 

We used an algorithm, described in \cite{p3Ref1}, to compute the probability
distributions of distance modulus shifts. The mean dispersion of the
magnitude shift due to gravitational lensing due to voids and halos
is $\sim3$ times larger than due to voids alone with a shell thickness,
\cite{p3Ref1}; the dispersion $\sigma_{m}$ due to 35 Mpc voids and halos
for sources at $z_{{\rm s}}=1.5$ is $\sigma_{m}=0.065-0.072$ (depending
on the model). The mean magnitude shift due to voids and halos
is of order $\delta m=-0.0010$ to $-0.0013$ (depending on the model). These values of $\sigma_m$ imply that large scale structure must be
accounted for in using luminosity distance determinations for 
estimating precise values of cosmological parameters, such as those
characterizing the dark energy equation of state.

We studied the distribution of magnitude shifts for three different
source redshifts in each of our models. The qualitative dependence
on redshift is similar to that of the previous void-only models \cite{p3Ref1}.
We find that the voids do not significantly change the variance but
do significantly change the demagnification tail and the mode.
The mode lies on the demagnification side and the variance is largely due to halo intersections. The scale
of voids is unimportant and the only discernable effect in the mode
is seen when the void interior is smoothly distributed matter. As
a result, our models bracket the range of possibilities of magnifications.

Our simple and easily tunable model for void and galaxy halo lensing
can be used as a starting point to study more complicated effects.
For example, one can use various algorithms to generate realizations
of distributions of non-overlapping spheres in three dimensional space.
Given such a realization one could use the algorithm of this paper
to study correlations between magnifications along rays with small
angular separations, which would be relevant to future small beam
surveys \cite{p3Ref23}. Finally, our model is in general agreement with other simplified
lensing models in the literature that focus on lensing due to both
halos and larger scale structures, \cite{p3Ref12, p3Ref14, p3Ref15, p3Ref16, p3Ref17, p3Ref18, p3Ref19, p3Ref20, p3Ref21, p3Ref22}.

Our results for $\sigma_m$ in the SRN void model
are represented within about 20\% by an analytic
model presented in detail in Appendix B. This model ascribes the
magnitude shift entirely to the fluctuations in light beam convergence
that results from passage through underdense cores and overdense halos;
thus it ignores the contribution from shear, which we have found to
be relatively small empirically.
The final result for $\sigma_m^2$ is a sum of these two contributions.
[See Eqs. (B20) and (B22)]. Although our simulations assumed a single
halo mass $M_h$, radius $R_h$ and concentration $C$, and a single void 
radius, the analytic model allows distributions for these key quantities. 
The contribution from halos is proportional to a suitably weighted 
mean of $M_h\Psi_2(C_h)/R_h^2$, where $\Psi_2(C_h)$ is defined in
Eq. (B16) and is displayed in Fig. 13. The contribution from the 
underdense void cores is proportional to the mean void radius. Typically,
the contribution to $\sigma_m^2$ from halos is much larger than the
contribution from void cores so $\sigma_m^2$ is larger for more massive
or more compact halos.

We have seen that the results of our simulations depend on whether the
underdense core consists of smoothly distributed dark matter (SRN)
or is itself clumped into halos (SRR). In the extreme case in which
the underdense core is entirely made of halos, the results do not
depend on the core density, and is equivalent to the SR model 
that consists of halos distributed randomly within a void. 

Although the analytic model was only developed for the SRN model,
it could also be applied to the SRR model with any prescription
for the fraction of the mass of the underdense core that is clumped
into halos. For example, in the extreme case of total clumping
we can use the analytic models in Appendix B with the substitution 
$f=1$ for all $z$ in all expressions derived there. For intermediate
cases, a prescription for the fraction of the underdense core that
remains smooth rather than clumped would be needed.

\begin{acknowledgments}
This research was supported at Cornell by NSF grants PHY-0968820 and
PHY-1068541 and by NASA grant NNX11AI95G.
\end{acknowledgments}

\section*{APPENDIX A: COLUMN DEPTH}

In this appendix, we describe how we calculate the column density
encountered by rays passing through halos in the SRN model. In our
model for the voids, we break up the bounding shell of mass into halos
with mass $M_{{\rm halo}}$. As a light ray passes through one of
these halos, the beam will acquire some integrated column depth, $\eta=\int\rho\left(z\right)dz$,
where the random variable $\eta$ depends on the impact parameter
of the beam with respect to the halo center, $\rho\left(z\right)$
is the density profile of the halo and $y$ is the physical coordinate.
The maximum value is for a beam going right through the center, and
diverges for our NFW profile \ref{3p2.1}. 

We use an NFW profile \ref{3p2.1}, for the matter distribution in halos,
and the column depth is 
\begin{equation}
\eta\left(b\right)=(2\rho_{0}R_{{\rm s}}^{3})\int_{0}^{\sqrt{\left(CR_{{\rm s}}\right)^{2}-b^{2}}}\frac{dz}{\sqrt{z^{2}+b^{2}}\left(\sqrt{z^{2}+b^{2}}+R_{{\rm s}}\right)^{2}}\tag{{A1}}\label{3pA1}
\end{equation}
where we have changed from radial (in Eq. (\ref{3p2.1})) to Cartesian coordinates.
Here the physical impact parameter is $b$ and $C$
is the ratio of the radius of the halo to its core radius. This reduces
to the sum of two contributions
\begin{equation}
\eta_{{\rm halo}}\left(b\right)=\eta_{{\rm core}}\Theta\left(R_{{\rm s}}-b\right)+\eta_{{\rm out}}\Theta\left(b-R_{{\rm s}}\right)\Theta\left(CR_{{\rm s}}-b\right)\tag{{A2}}\label{3pA2}
\end{equation}
where the relationship between the column depth $\eta\left(b\right)$ and the corresponding
lensing convergence, $\kappa\left(b\right)$, is, from Eq. (3.1), 
\[
\eta\left(b\right)=\left(\frac{4\pi Ga_{{\rm ex}}\left(z\right)}{c^{2}}\frac{y\left(y_{{\rm S}}-y\right)}{y_{{\rm S}}}\right)^{-1}\kappa\left(b\right).\tag{{A3}}\label{3pA3}
\]
The lensing convergences is listed in Eqs. (\ref{3p3.4}) and (\ref{3p3.5}). Again,
the contribution from outside a radius of $CR_{{\rm s}}$ is zero.
The mean column density of halos is defined as 
\[
\eta_{{\rm halo}}=\frac{M_{{\rm halo}}}{\pi\left(CR_{{\rm s}}\right)^{2}}.\tag{{A4}}\label{3pA4}
\]

For the halos, even more important that $\eta$ is $d\eta/d\alpha$,
where $\alpha=b^{2}/R_{{\rm s}}^{2}$; this is because the probability
distribution for $\alpha$ is $d\alpha/C^{2}$, and therefore the
probability distribution for $\eta$ for a single halo is 
\[
P_{{\rm halo}}\left(\eta\right)=\frac{1}{C^{2}\left|d\eta_{{\rm halo}}/d\alpha\right|}.\tag{{A5}}\label{3pA5}
\]
This is the quantity plotted in Figure 1.

\section*{APPENDIX B: ANALYTIC ESTIMATE OF STANDARD DEVIATION}

In this appendix, we derive analytic results for the standard deviation of magnifications. Consider the mean of the contribution to the convergence (3.1) from the underdense core of the $j$-th void, which we will denote by $\kappa_{c,j}$. We find 
\[
\kappa_{c,j}=-{3H_0^2\Omega_M\over 2}\times{f_jy_j(y_S-y_j)\over a_jy_S}
\times 2\sqrt{Y_j^2-p_j^2}\tag{{B1}}
\]
where $Y_j$ is the comoving radius of the void, $a_j$ is the scale factor and $f_j$ is the fraction of the total void mass on the surface at redshift $z_j$. After averaging over impact paramters we obtain
\[
\langle\kappa_{c,j}\rangle=-2\Omega_MH_0Y_j{f_jH_0y_j(y_S-y_j)\over a_jy_S},\tag{{B2}}
\]
and therefore the net expected convergence from voids is 
\[
\langle\kappa_c\rangle=\sum_j\langle\kappa_{c,j}\rangle=-2\Omega_MH_0
\sum_j{f_jY_jH_0y_j(y_S-y_j)\over a_jy_S}~.\tag{{B3}}
\label{kappac}
\]
Assuming a typical radius $Y_j \sim Y_{\rm{void}}$, 
there are about $(H_0Y_{\rm{void}})^{-1}$ terms in the sum, and consequently the overall average 
is $\sim\Omega_M$. In the limit that there are many voids along a
given line of sight, we can replace the sum by an integral. The number of
voids per interval $dy$ of comoving distance is $dy/2R$, and therefore
if we define $\xi=H_0y$ we find
\[
\langle\kappa_c\rangle=-2\Omega_MH_0\sum_j{f_jR_iH_0y_j(y_S-y_j)\over a_jy_S}
\]
\[
\to-\Omega_M\int_0^{\xi_S}{d\xi f(\xi)\xi(\xi_S-\xi)\over a(\xi)\xi_S}~,\tag{{B4}}
\label{kapcint}
\]
where $\xi_S=H_0y_S$. Equation (B4) does not depend on any void properties (apart from the value of $f$ today) and remains
valid if there is a distribution of void sizes, for example.

On average, the contribution $\kappa_h$ from halos to the lensing convergence must cancel the contribution (B4) from voids, i.e., $\left<\kappa_h\right>=-\left<\kappa_c\right>$. 

We assume statistical independence of halos within voids from the core (i.e., true if halo radii are small) and also of voids from one another. The overall variance is a sum of individual halo and core variances. As in our simulations, we neglect clustering of halos, which would
introduce correlations among them, and assume that dark matter is
confined to the halos and underdense core. These assumptions could
be relaxed in a more sophisticated model. 
We see that
\[
\langle\kappa_c^2\rangle-\langle\kappa_c\rangle^2
=\sum_j\langle\kappa_{c,j}^2\rangle-\langle\kappa_{c,j}\rangle^2
\]
\[
={1\over 2}\Omega_M^2H_0^2\sum_j{Y_j^2f_j^2H_0^2y_j^2(y_S-y_j)^2\over a_j^2y_S^2}\tag{{B5}}
\]
since the averages for $i\neq j$ vanish. The sum is $\lesssim(H_0Y)^{-1}$ and therefore 
$\langle\kappa_c^2\rangle\lesssim\Omega_M^2H_0Y$.

Let us now consider halos residing in void $j$. We include the possibility that there are different
types of halo with different properties, and label the types by $\alpha$.
For a given halo $i$ of type $\alpha$ passed through by the line of sight at physical impact parameter
$\baj$ relative to its center, the contribution to $\kappa_h$ is
\[
\kai={8\pi G a_j y_j(y_S-y_j)\over y_S}\int_{\baj}^{R_{h,\alpha}}{dr\,r\,
\rho_\alpha(r)\over\sqrt{r^2-\baj^2}}\tag{{B6}}
\]
where $R_{h,\alpha}$ is the physical halo radius, and 
$\rho_\alpha(r)$ is the physical density within the halo. The average over impact parameters $\baj$ is
\[
{2\over R_{h,\alpha}^2}\int_0^{R_{h,\alpha}}d\baj \baj\int_{\baj}^{R_{h,\alpha}}
{dr\,r\,\rho_\alpha(r)\over\sqrt{r^2-\baj^2}}
\]
\[
={M_{h,\alpha}\over 2\pi R_{h,\alpha}^2}~,\tag{{B7}}
\]
where $M_{h,\alpha}$ is the total halo mass.
Therefore the average over impact parameters through a given halo is
\[
\langle\kai\rangle={4GM_{h,\alpha}a_j y_j(y_S-y_j)\over R_{h,\alpha}^2y_S}~.\tag{{B8}}
\]
If all of the halos reside in the voids (i.e., none in the FRW exterior), then the expected number of
intersections of the light path with a halo is $\nai R_{h,\alpha}^2/a_j^2Y_j^2$, where $\nai$
is the expected number of halos of ``type $\alpha$'' in the void, and we take account of the fact that 
$R_{h,\alpha}$ is
a physical radius. Then we get a total contribution from halos per mass-compensated void equal to
\[
\sum_\alpha {\nai R_{h,\alpha}^2\langle\kai\rangle\over a_j^2Y_j^2}
={4Gy_j(y_S-y_j)\over a_jY_j^2}\sum_\alpha \nai M_{h,\alpha}~;\tag{{B9}}
\]
the sum is the total mass in halos, which must compensate the underdensity, so
\[
\sum_\alpha \nai M_{h,\alpha}={f_jH_0^2\Omega_MY_j^3\over 2G}\tag{{B10}}
\]
and therefore
\[
\sum_\alpha {\nai R_{h,\alpha}^2\langle\kappa_{h,\alpha}\rangle\over a_j^2Y_j^2}
={2f_jH_0^2\Omega_MY_jy_j(y_S-y_j)\over a_jy_S}
=-\langle\kappa_{c,j}\rangle~.\tag{{B11}}
\label{kappamean}
\]
Therefore the average per mass-compensated void cancels as expected.
This cancellation is actually independent of the distribution of halos 
within the void but depends on the assumption that all halos are associated
with voids.

Let us suppose that
\[
\kai={4GM_{h,\alpha}a_\alpha y_j(y_S-y_j)\over R_{h,\alpha}^2y_S}\,F_\alpha
\left({\baj^2C_{h,\alpha}^2\over R_{h,\alpha}^2}\right)\tag{{B12}}
\label{kappahamodel}
\]
where $C_{h,\alpha}$ is dimensionless. This form assumes that the density profile for halo type $\alpha$ has one scale
parameter, $R_{h,\alpha}/C_{h,\alpha}$, although it does not necessarily assume that the density profiles are
the same for all halos either in form or in the parameter $C_{h,\alpha}$.
We do assume that
\[
\int_0^1 dx\,F_\alpha(xC^2_{h,\alpha})=1
={1\over\Cha^2}\int_0^{\Cha^2}dq\,F_\alpha(q),\tag{{B13}}
\label{normalization}
\]
independent of the value of $C_{h,\alpha}$. With this normalization,
if we let $\rho_\alpha(r)=\rho_{0,\alpha}
\rhohat_\alpha(rC_{h,\alpha}/R_{h,\alpha})$
we find
\[
F_\alpha\left({\baj^2C_{h,\alpha}^2\over R_{h,\alpha}^2}\right)
\]
\[
\;\;\;\;\;\;\;\;\;={2\pi \rho_{0,\alpha}R_{h,\alpha}^3\over M_{h,\alpha}C_{h,\alpha}}
\int_{\baj\Cha/\Rha}^{C_{h,\alpha}}{du\,u\,\rhohat_\alpha(u)\over\sqrt{u^2-
\baj^2\Cha^2/\Rha^2}}~.\tag{{B14}}
\label{Fdef}
\]

Equation (\ref{kappahamodel}) with the normalization (\ref{normalization})
leads to Eq. (\ref{kappamean}) as expected, but it also implies contributions
from each halo to the variance given by
\[
\kai^2-\langle\kai\rangle^2=\left[{4GM_{h,\alpha}a_\alpha y_j
(y_S-y_\alpha)\over R_{h,\alpha}^2y_S}\right]^2
\]
\[
\times\left\{\left[F_\alpha\left({\baj^2C_{h,\alpha}^2\over R_{h,\alpha}^2}\right)\right]^2-1
\right\}~.\tag{{B15}}
\]
Averaging over impact parameters implies 
\[
\langle\kai^2\rangle-\langle\kai\rangle^2
\]
\[=\left[{4GM_{h,\alpha}a_\alpha 
y_j(y_S-y_\alpha)\over R_{h,\alpha}^2y_S}\right]^2
\left[\int_0^1 dx\,F^2(x\Cha^2)-1\right]
\nonumber\\
\]
\[
\;\;\;=\left[{4GM_{h,\alpha}a_\alpha y_j
(y_S-y_\alpha)\over R_{h,\alpha}^2y_S}\right]^2
\Psi_2(\Cha)~.\tag{{B16}}
\label{variance}
\]
where the function $\Psi_2(\Cha)$ for NFW profiles is plotted in Figure 13.

\begin{figure}
\includegraphics[scale=0.59]{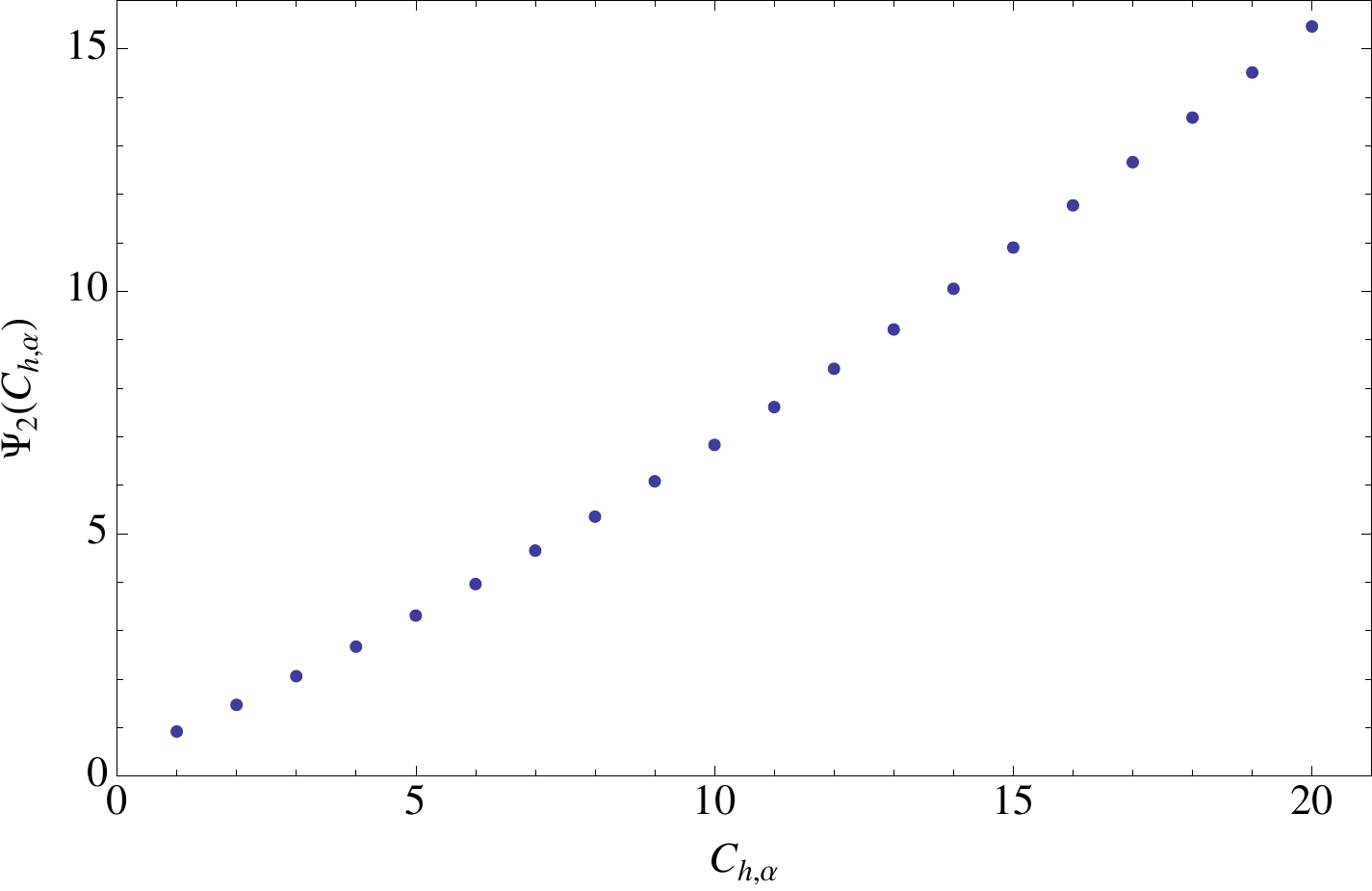}

\caption{Plot of the behavior of $\Psi_2(\Cha)$ as a function of the concentration parameter $\Cha$ of NFW halos.}
\end{figure}

We can use these results to determine the expected contribution of halos
in a given mass-compensated void to the variance. %For this purpose, let
%us assume that we have averaged over impact parameters and that $\alpha$
%refers to a class of halos characterized by $(\Mha,\Rha,\Cha)$.
Since total mass in halos in void $j$ is $f_j\Omega_MH_0^2R_j^3/2G$, if the
fraction of this total in halos of type 
$\alpha$ is $\eali$, then expected number of type $\alpha$
is 
\begin{equation}
\nai={\eali f_j\Omega_MH_0^2Y_j^3\over 2G\Mha}.
\tag{{B17}}
\label{naieqn}
\end{equation}
The expected number of intersections
with type $\alpha$ is
\[
\nuai
={\eali f_j\Omega_MH_0^2Y_j\Rha^2\over 2a_j^2G\Mha}.
\tag{{B18}}
\label{nuaieqn}
\nonumber\\
\]
The contribution from halos in mass-compensated void $j$ to the total variance 
is therefore
\[
{8f_jG\Omega_MH_0^2Y_jy_j^2(y_S-y_j)^2\over y_S^2}\sum_\alpha{\eali\Mha\Psi_2(\Cha)
\over\Rha^2}~.\tag{{B19}}
\]
Summing over all mass-compensated voids out to the source we get 
\[
\sigma^2_\kappa=\Omega_M\sum_j\left[8\sum_\alpha{\eali G\Mha\Psi_2(\Cha)
\over\Rha^2}+{f_j\Omega_MH_0^2R_i\over 2a_j^2}\right]
\]
\[
\times{H_0^2Y_jf_jy_j^2(y_S-y_j)^2\over y_S^2}~.
\tag{{B20}}
\label{sigmakappa}
\]
Equation (\ref{sigmakappa}) shows that the overall variance depends on halo 
properties primarily via the typical value of $G\Mha/\Rha^2$, the gravitational
acceleration characteristic of the outer regions of the halo. When this is
large compared with the mean gravitational acceleration of the void as a whole,
$f_j\Omega_MH_0^2R_i/a_j^2$, the halos dominate the dispersion.
There is also a hefty numerical factor $8\Psi_2(\Cha)\sim 50$ favoring
the contribution from halos.

If $\eali=\eal$ is actually independent of $i$ then we can factor out the sum
over $\alpha$ in the first term of Eq. (\ref{sigmakappa}): define
a dimensionless parameter
\[
\gbarh\equiv {16\over H_0}\sum_\alpha{\eal G\Mha\Psi_2(\Cha)\over\Rha^2}
\]
\[
\;\;\;\;\;\;\;\;\;\;\;={0.175\langle\Mha\Psi_2(\Cha)/\Rha^2\rangle\over
h\times 10^{12}\Msun/(300\,{\rm kpc})^2\times 6.84.}
\tag{{B21}}
\label{gbarhdef}
\nonumber\\
\]

We rewrite Eq. (\ref{sigmakappa}) as
\[
\sigma^2_\kappa={\Omega_M\over 2}\gbarh\sum_j
{(H_0Y_jf_j)H_0^2y_j^2(y_S-y_j)^2\over y_S^2}
\]
\[
\;\;\;\;\;\;\;\;\;\;\;\;\;\;\;\;\;+{\Omega_M^2\over 2}\sum_j{(H_0Y_jf_j)^2H_0^2y_j^2(y_S-y_j)^2\over a_j^2y_S^2}.
\tag{{B22}}
\label{newsigkap}
\]
The first term in Eq.(\ref{newsigkap}) is $\sim\gbarh$ and the second
is $\sim \Omega_MH_0\langle Y_j\rangle\approx 3.5\times 10^{-3}h
\langle Y_j\rangle/35\,{\rm Mpc}$. Eq. (\ref{gbarhdef}) 
suggests that halos dominate.

Now all of the parameters in Eq. (\ref{newsigkap}) can be varied over
distributions. To keep things as simple as possible, let us assume that
$f_j$ is simply a function of redshift; that is, neglect the possible
dependence of $f$ on the size of mass compensated voids. As a further
simplification, we turn the sums into integrals. If we had a single
void radius $R$ the sums would turn into integrals by noting that 
there are $dy/2R$ voids per range $dy$; for a distribution of void
sizes we can use this substitution with $R_i\to\langle R\rangle$ in
the second sum. If we also define $\xi=H_0y$ then with these
simplifications
\[
\sigma^2_\kappa(\xi_S)={\Omega_M\over 4}\gbarh
\int_0^{\xi_S}{d\xi\,f(\xi)\xi^2(\xi_S-\xi)^2\over \xi_S^2}
\]
\[
\;\;\;\;\;\;\;\;\;\;\;\;\;\;\;\;\;\;\;+{\Omega_M^2\over 4}H_0\langle R\rangle\int_0^{\xi_S}{d\xi\,f^2(\xi)\xi^2(\xi_S-\xi)^2
\over a^2(\xi)\xi_S^2}
\tag{{B23}}
\label{sigcompute}
\]
The relationship between the variance of $\kappa$ and the variance of magnifications $\sigma_m$, for small deviations, is approximated by 
\[
\sigma_m^2=\left({5\over\log\left(10\right)}\right)^2
\sigma_\kappa^2.
\tag{{B24}}
\]

\end{document}